\documentclass[graybox]{svmult}

\usepackage{type1cm}        % activate if the above 3 fonts are
\usepackage{makeidx}         % allows index generation
\usepackage{graphicx}        % standard LaTeX graphics tool
\usepackage{multicol}        % used for the two-column index
\usepackage[bottom]{footmisc}% places footnotes at page bottom

\usepackage{newtxtext}       % cf

\usepackage{newtxmath}       % selects Times Roman as basic font

\newcommand\actaa{{Acta Astron.}~}%   

\newcommand\araa{{ARA\&A}}%   
\newcommand\apj{{ApJ}~}%   
\newcommand\apjl{{ApJL}~}%   
\newcommand\apjs{{ApJS}~}%   
\newcommand\aap{{A\&A}~}%   
\newcommand\mnras{{MNRAS}~}%   
\newcommand\pasa{{PASA}~}%   
\newcommand\prl{{Physical Review Letters}~}%   
 \newcommand\memsai{{Memoirs Soc. Astr. It.}~}%   
\newcommand\nat{{Nature}~}%   
\def\simgt{\lower.5ex\hbox{$\; \buildrel > \over \sim \;$}}
\def\simlt{\lower.5ex\hbox{$\; \buildrel < \over \sim \;$}}

\newcommand\teff{T$_{\rm eff}$}

\newcommand{\msun}{\ensuremath{\, {M}_\odot}}
\newcommand{\Msun}{\ensuremath{\, {M}_\odot}}

\newcommand{\Mc}{M$_{\rm c}$}
\newcommand{\Min}{M$_{\rm in}$}

\newcommand{\Porb}{$P_{\rm orb}$}
\newcommand{\Pmin}{$P_{\rm min}$}
\newcommand{\Mdon}{$M_{\rm d}$}

\newcommand{\Mdot}{$\dot{M}$}
\newcommand{\Pdot}{${\dot{P}}_{\rm orb}$}
\newcommand{\Pdotspin}{${\dot{P}}_{\rm s}$}
\newcommand{\Pspin}{${P_{\rm s}}$}
\newcommand{\Psh}{${P_{\rm sh}}$}
\newcommand{\Pbif}{${P_{\rm bif}}$}

\newcommand{\Bs}{$B_{\rm s}$}
\newcommand{\tauconv}{$\tau_{\rm conv}$}

\makeindex             % used for the subject index
\begin{document}

\title*{Origin and binary evolution of millisecond pulsars}
\author{Francesca D'Antona and Marco Tailo}
\institute{Francesca D'Antona \at INAF-Rome Observatory, Address of Institute, \email{francesca.dantona@inaf.it}
\and Marco Tailo \at University of Padua, Address of Institute \email{marco.tailo@gmail.com}}
\maketitle
%\abstract*{
\abstract{We summarize the channels formation of neutron stars (NS) in single or binary evolution and the classic recycling scenario by which mass accretion by a donor companion accelerates old NS to millisecond pulsars  (MSP). We consider the possible explanations and requirements for the high frequency of the MSP  population in Globular Clusters.
Basics of binary evolution 
%(Sect.\,2) 
are given, and the key concepts of systemic angular momentum losses are first discussed in the framework of the secular evolution of  Cataclysmic Binaries.
%, their 2-3\,hr period gap and minimum period. %(Sect.\,3). 
%MSP binaries with compact companions of mass $M_{\rm c}$ ---another NS, or a Helium, Carbon--Oxygen, or Oxygen--Neon WD---  reach their present location in the \Porb\ versus $M_{\rm c}$\ plane as result of  previous evolution. %(Sect.\,4).  
MSP binaries with compact companions represent end-points of previous evolution.
In the class of systems characterized by short orbital period \Porb\ and low companion mass %$M_{\rm d}$\  
we may instead be catching the recycling phase `in the act'. These systems are in fact 
either MSP, or low mass X--ray binaries (LMXB), some of which accreting X--ray MSP (AMXP), or even `transitional' systems from the accreting to the radio MSP stage.
The donor structure is affected by irradiation due to X--rays from the accreting NS, or by the high fraction of MSP rotational energy loss emitted in the $\gamma$\ rays range of the energy spectrum.  X--ray irradiation leads to cyclic LMXB stages, causing super--Eddington mass transfer rates during the first phases of the companion evolution, and, possibly coupled with the angular momentum carried away by the non--accreted matter, helps to
explain the high positive \Pdot's of some LMXB systems  
% the large differences in \Pdot\ among different systems, 
and account for the (apparently) different birthrates of  LMXB and MSP.  Irradiation by the MSP may be able to drive the donor to a stage in which either radio-ejection (in the redbacks) or mass loss due to the companion expansion, and `evaporation' may govern the evolution to the black widow stage and to the final disruption of the companion.      }

\section{Introduction}
Understanding millisecond pulsars (MSP) has a particular appeal for stellar evolution, for its breadth among a number of still open problems, first of all those belonging to the neutron star (NS) formation, such as the different paths to supernova explosions, the  supernova kicks and the initial mass distribution of NSs. In addition, the binary evolution to the MSP stage is either affected by crude parametrization of difficult phases (common envelope, systemic angular momentum losses) or by subtle issues, such as the description of `illumination' by X--rays or by the MSP energy loss, and its effect on the structure of the companion star in the short and long term evolution. Thus this review focuses on a few problems and ignores most of the surrounding rich physics of these objects.\\
We shortly discuss the channels of neutron star (NS) formation in single and binary stellar evolution and deal with the problem of the high frequency of MSP in Globular Clusters (GCs) in Sect.\,\ref{sec:2}. Then we summarize basic concepts of binary evolution in Sect.\,\ref{sec:3}. As a comparison key study, we discuss the secular evolution of Cataclysmic Binaries (CBs) in Sect.\,\ref{sec:4}, and give a coarse comparison between the orbital period distribution of these systems and that of MSP and low mass X-ray binary systems. Sect.\,\ref{sec:5}
 discusses the evolution of different classes of MSP binaries, those having compact remnant companions, while Sect.\,\ref{secular} deals with the evolution of low mass -- short period LMXB and MSP. The role of `irradiation' or `illumination' is discussed and found relevant, together with `radioejection' and `evaporation' to explain the different period distribution of the binary MSP. 

\section{The origin of millisecond pulsars}
\label{sec:2}
The topic consists of two generally independent problems: the formation of neutron stars (NS) and the acceleration of the NS to millisecond periods. The fact that MSPs are statistically much more abundant in Globular Clusters points to an important formation role for these stellar ensembles.
%Actually, in some formation channels, the formation of the MSP is concomitant with the formation of the NS, but this is not the rule.
%%% FIGURE 1 %%%%%% (fig1)
\begin{figure}[t]
\includegraphics[scale=0.17]{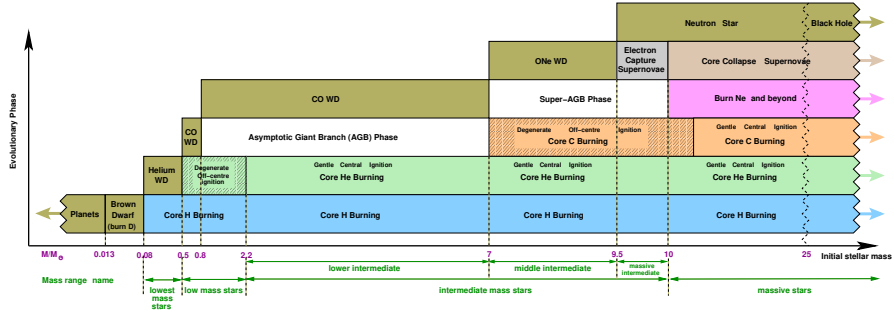}
\caption{From Karakas \& Lattanzio \cite{karakas2014}, the scheme shows how the initial stellar mass determines the main nuclear burning phases, as well as the fate of the final remnant. NS are the remnants of ecSN and CCSN, from $M\sim 9.5-25$\Msun.
%The mass limits are rough estimates, and depend on composition (here solar metallicity) in addition to details of the input micro- and macro-physics. 
}
\label{fig1}       % Give a unique label
\end{figure}

\subsection{The formation of single neutron stars}
The evolution of single stars as a function of their initial mass is schematically shown in Figure\,\ref{fig1}, from the work by Karakas \& Lattanzio (2014, \cite{karakas2014}) which we take as general reference for this Section. The mass boundaries between different kinds of evolutionary paths must be regarded as indicative only, because the precise mass values depends both on physical inputs (e.g. the metallicity) and on the assumptions made on those inputs of evolution which are not known from first principles (e.g. mass loss or core overshooting) and thus have to be educately parametrized.\\
Masses which ignite helium in the core within the age of the Universe ($\simgt$0.8\Msun) and up to $\sim$7\Msun\ develop carbon--oxygen (C-O) cores. Their evolution procedes along the Asymptotic Giant Branch (AGB) phase, through thermal pulses (\cite{iben1975}) and mass loss. Finally they become C-O white dwarfs (WDs). In a small range of masses above 7\Msun, which develop C-O core masses 1.1$\lesssim M_{C-O}/M_\odot \lesssim$1.35,  carbon is ignited in conditions of semi--degeneracy, the energy released is able to expand the core, which is further processed to become an oxygen--Neon (O-Ne) core. Above the core, the star experiences thermal pulses and mass-loss (`super--AGB' evolution). The final outcome is generally an O-Ne WD. If the core can grow beyond the Chandrasekhar limit, electron captures occur before the core can ignite Neon. The depletion of electron pressure
in the core induces the core collapse to nuclear densities in the event named ``electron-capture supernova" (ecSN)\footnote{In Fig.\,\ref{fig1} ecSN occur for an initial mass range of $\sim$0.5\Msun, but this value is very dependent on the modelization of the evolution of super--AGB phase \cite{poelarends2008}.}. In larger masses, the supernova explosion occurs when the star has ignited all elements up to iron, and the core collapse is caused by the iron ignition (core collapse supernova, CCSN). Up to an initial mass estimated to be $\sim$20-25\Msun, the remnant of CCSNs are neutron stars, while above a black hole is formed. \\
Thus NS can be born from either ecSN or CCSN, from an initial  mass range $\sim$9.5--25\msun.
%Although it may seem a detail, the channel of formation (ecSN or CCSN) plays an important role, especially when we are discussing the formation and evolution of millisecond pulsars (MSP), for which the location in a Globular Cluster (GC) environment results to be an important feature. 
\subsection{Formation of neutron stars in binaries}
\label{sec:22}
The scheme of neutron star formation is largely modified when we deal with the evolution of binary systems. In general, the mass range of both CCSN and ecSN becomes larger. 
\begin{enumerate}%[CCSN]
\item[1]{{\bf CCSN}: The binary channel to CCSN also includes the so-called delayed Type II SN, exploding in binaries in which the mass necessary for the event is obtained by mass exchange \cite{zapartas2017}. }
\item[2]{{\bf ecSN}: The binary evolution has an important role which can end in a higher frequency of ecSN events, as described by Podsiadlowski et al. \cite{podsi2004}. In binaries beginning mass transfer when the donor is evolving through the Herzsprung gap, before reaching the giant branch location, the star may avoid the second dredge up (2DU) phase which in single stars reduces the mass of the helium core remnant from the main sequence evolution. This He--core mass is the dominant factor in the following evolution, as it is converted into the initial C-O core mass by helium burning. If, for a given initial mass, the C-O core is larger, the larger is its probability to become an ecSN instead of a plain O--Ne WD.  }
\item[3] {\bf AIC ecSN}: in an interacting binary containing an O--Ne WD, accretion may push the WD above the Chandrasekhar mass, in an ecSN event named ``accretion induced collapse" (AIC) (e.g. \cite{nomoto1987}, \cite{freire-tauris2014}). %\cite{ablimit2015aic}).
\end{enumerate}
%Further cases of binary evolutions leading to NS formation are possible, and may explain the peculiarity of some pulsar binaries \cite{tauris-sennels2000}.

\subsection{The MSP formation}

We recall that timing measurements allow to detect the spin period of pulsars (\Pspin) and its time derivative (\Pdotspin). The hypothesis that the loss of rotational energy of the pulsar equals the amount of magnetic dipole radiation emitted yields an estimate of the average strength of the dipole component of the pulsar magnetic field at the surface (\Bs)
%%%% tolta formula per risparmiare spazio
%\footnote{in this description \Bs is the equatorial \textit{B}--field strength. An expression for the spin-down torque different than the magnetic vacuum dipole model can be found in \cite{spitkovski2006} and is adopted in \cite{tauris2012}. As we are now dealing with a qualitative description, here we use the classic formulation given in \cite{batta1991}, where the approximation implicit in this expression is discussed.}: 
%\begin{equation}
%B_{\rm s} = \left( {{3c^3I}\over{8 \pi^2 R^6}} P_{\rm s}\dot{P}_{\rm s} \right)^{1/2} \times {1 \over \sin\alpha}\simeq 3.2\times 10^{19} (P_{\rm s} \dot{P}_{\rm s})^{1/2} ~{\rm G}
%\end{equation}
%where $c$\ is the speed of light, $I$\ is the moment of inertia of the NS and $R$\ is its radius,  and $\alpha$\ is the angle between the magnetic field and the spin axis $0 < \alpha \leq 90^{\circ}$. Standard values ($I=10^{45}$\,g cm$^2$, $R=10$\,Km and $P$\ in seconds) are used for the evaluation given at the right. \\
A spin-down age may also be defined by 
%\begin{equation}
$\tau = {P_{\rm s} \over {2 \dot P_{\rm s}}}$ \\
%\end{equation}
The \Pspin\ and \Pdotspin\ data then allow to build the \Bs\ versus \Pspin\ diagram, which best describes the evolution of pulsars. %(SEE FIGURE XX IN CHAPTER XY). 
%The majority of pulsars occupy a region around \Pspin$\sim$1\,s and \Bs$\sim 10^{11}-10^{13}$G. These are recognized as the `young' pulsars, whose spin has been achieved at the formation of the NS. 
  %\begin{svgraybox}
%In the \Bs\ versus \Pspin\ diagram 
%\end{svgraybox}
%On the contrary, 
MSP are identified as a class of `old' NSs. The very low \Pdotspin's ($< 10^{-18}$), a modest loss of rotational energy due to emission of magnetodipole waves,  imply that their magnetic field is 3--5 orders of magnitude lower than that of young pulsar.  Thus these pulsars have high $\tau$\ and are visible as MSP for gigayears after their formation. 
%FIRST THOUGHTS: decays spontaneously on a e- folding timescale t?  107 ? 108 yr (e.g. Lyne, Anderson & Salter 1982) from an initial value  1031 G cm3 to a final value  1026 ? 1027 G cm3, when the decay probably stops (e.g. Bhattacharya & Srinivasan 1986)
%%% tolto per preprint
%%%IN CHAPTER XX (PAPITTO) AND CHAPTER XY (DEEPTO) the  mechanisms which may be at the basis of the low magnetic field in MSP are reviewed.
%%%
%Alessandro revised
%The magnetic field evolution of MSPs has been considered in different contexts (see, e.g. \cite{taamvdh1986}). The field observed  is low either because of long--term ohmic decay of the crustal field during the pre--accretion lifetime of the NS, or by rapid ohmic decay in the accretion-heated crust \cite{konar-batta1997}, or by accretion--induced magnetic field screening \cite{cumming2001}. As far as we know, no detailed evolution taking into account the accretion influence on the magnetic field change has been performed, although suggested, e.g. in \cite{taurisvdh2006}.
%because the surface field has been buried during the first phases of mass accretion \cite{taamvdh1986}.
\\
Soon after the discovery of the first MSP, PSR 1937+21 (a single NS, but possibly previously host in a binary \cite{backer1982}), a model was proposed to explain its 1.55\,ms spin period as due to accretion of mass and angular momentum from a binary companion. PSR\,1937+21 was then recognized as a ``recycled"\footnote{This term had already been used to describe the status of the pulsar in the Hulse-Taylor binary NS PSR B1913+16, discovered in 1974} pulsar\cite{alpar1982}. 
%\cite{rada1982,alpar1982}.
The recycling model was also testified \cite{fabian1983} by the existence of accreting NS, in the Low Mass X--ray Binaries (LMXBs), and confirmed by the discovery of accreting X-ray MSPs (AMXPs), whose first example has been  SAX\,J1808-3658\cite{wijnands1998}. 
Following  \cite{illarionovsunayev1975}, accretion can occur on the NS if the magnetospheric radius (where the accretion disc is destroyed by the magnetic pressure) 
%the accretion disk is truncated because of one of the followingreasons: (i) the interaction with the magnetic field of the NS, which truncates the disc at the magnetospheric radius RM, at which the accretion flow is channeled along the magnetic field lines towards the magnetic poles onto the NS surface; (ii) the presence of the NS surface itself at R; and (iii) the lack of closed Keplerian orbits for radii smaller than the marginally stable orbit radius, RMSO (at few ? depending on the mass and spin of the compact object ? Schwarzschild radii from the NS centre). The position of RM is determined by the istantaneous balance of the pressure exerted by the accretion disc and the pressure exerted by the NS magnetic field:
is inside the corotation radius, where the keplerian velocity equals the rotation velocity of the NS. 
Mass and angular momentum are then accreted from the inner rim of the disk, and the NS spins up, so the corotation radius becomes closer to the NS. 
When, the corotation radius  becomes smaller than the magnetospheric radius, matter is centrifugally dispersed and can not accrete (propeller effect). 
We may expect that in MSPs, accretion of mass and angular momentum on the NS through the accretion disc may have already   pushed the NS spin at the equilibrium value which is roughly equal to the keplerian angular frequency at the inner rim of the disc \cite{ghoshlamb1979}. 

\subsubsection{The lack of sub-millisecond pulsars and the radio--ejection}
\label{submsp}
The spin evolution of NS depends on important details, e.g. on the role of the neutron star matter equation of state and of the same NS spin in determining the NS radius, on how the magnetic field of the NS is modified by accretion, and on the disk description itself. 
%In detailed computations of accretion of mass and angular momentum, also the difference between the baryonic mass  and the gravitational mass must be considered \cite{lavagetto2004}. 
The minimum observed spin period of MSP (1.55\,ms) can be obtained by the modest mass accretion $\sim$0.1\msun, for any assumption on the equation of state for the neutron star matter, and is much longer that the limiting period \Psh\ below which the star becomes unstable to mass shedding at its equator (see  \cite{cook1994}). Further, a reasonable transfer of $\sim$0.3\msun\ appears sufficient to bring the NS down to its maximum spin \cite{burderi1999} well below 1\,ms, posing the problem of why we do not see sub-millisecond pulsars. 

Apart that observing such short periods is a challenging task, it is possible that the energetic MSPs do not allow further accretion. Nevertheless, the corotation radius is close to the NS surface, so a strong energy loss from the NS is needed to efficiently get rid, via the propeller mechanism, of the matter further lost by the donor during its whole secular evolution.  Burderi et al. \cite{burderi2001} proposed instead that a `radio-ejection' mechanism sets in when two circumstances happen together: the radio pulsar becomes active and the magnetic pressure due to the NS overcomes the disk internal pressure, so that it the disk becomes unstable and the MSP prevents further accretion directly at the inner lagrangian point. The specific angular momentum associated to this mass loss will be dominant to estimate the further secular evolution (see Sect.\,\ref{secular}). The radio-ejection conditions depend on a high power of the pulsar spin, but it is more efficient than the propeller, as the mass loss is driven by the system secular evolution and not by the rotational energy loss of the MSP. The role of radio-ejection (claimed \cite{burderi2002} for  PSR\,J1740--53) may be important and affect many MSP binaries. 
 
\subsection{Binary and single MSPs and the Globular Clusters environment}
\label{gc1}
The recycling model implies that there has been a stage of the NS life during which it has been a component of an interacting binary. The (few) single MSPs observed are generally regarded as previous binaries whose companion has been fully evaporated by the pulsar radiation, if it is the remnant of an LMXB. 
%In particular cases, a possible ``coalescence" model may apply also from the evolution of high mass X--ray binaries (HMXB), as shown in \cite{batta1991}.\\
Was the NS formed with the same companion which later on caused the recycling? This is possible in particular cases, but not strictly necessary for the bulk of MSPs found in Globular Clusters (GC), where binaries containing NS may be formed by tidal capture, binary exchanges or direct collisions\cite{ivanova2008}. MSPs in the field should then be the outcome of the evolution of pristine binary systems (see Sect\,\ref{DNS}, \ref{IMtoMSP}, \ref{mspcaseb})\footnote{Note anyway the scenario proposed by Grindlay et al. 1985 \cite{grindlayhertz1985}: also the bulge LMXBs may have been formed in GCs, later on destroyed by repeated tidal stripping and shocking in the galactic plane.}, while MSPs in GCs can  result both from the evolution of primordial systems and from new binaries formed after the NS formation by means of dynamical interactions \cite{king2003}. And in fact the dynamical role of the GC environment must be overwhelming, as $\sim$40\% of MSPs are found in GCs, despite the Galaxy is 10$^3$\ times more massive than the entire GC system\footnote{MSPs in the galactic field are 252 , and those in  GCs are 149 at the time of writing (Dec. 2019)}. A similar overabundance holds for LMXBs, reinforcing the hypothesis.
%In \cite{batta1991} we 

The dominant presence of MSPs and LMXBs in GCs poses the question that the GC must retain at least a fraction of the NS. This is not so obvious, as, in fact, we expect that most of the NS 
born from CCSN are expelled from clusters, as there is a momentum kick resulting from the supernova, due to the explosion asymmetries. The kick is a consequence of non radial hydrodynamic instabilities, such as neutrino driven convection or accretion shock instability (see \cite{janka2016} and references therein) in the collapsing stellar core.

Average observed kick velocities are very large, obviously much larger than the escape velocity from typical clusters. The important issue is then to look at the constraints on the fraction of {\it low velocity} pulsars.   
It is estimated that the average 3D initial pulsar velocity in the galactic disk is $\sim$400\,Km/s, with a 1D rms velocity of $\sigma=265$\,Km/s  \cite{hobbs2005} and in these conditions only a fraction as low as 3$\times 10^{-3}$ have space velocities below 60\,Km/s and  could be retained in GCs. 
%for a Maxwellian with $\sigma$ = 265\,Km/s of CCSN
%%%% revisione dopo Mapelli
In other models \cite{faucherkaspi2006} the fraction of pulsars below 60\,Km/s  varies in the ample range 0.012--0.135.
%Some authors interpret the pulsar's proper motion data as a two component distribution, one of which is peaked at relatively low velocity \cite{arzoumanian2002, verbunt2017}. The latter analysis implies that $\sim 5$\% of pulsars has space velocities below 60\,Km/s. 

The binary channels are favoured for the NS retention, as we expect that their SN are subject to smaller kicks\footnote{The dependence of the kick velocity on the mass ejected in the SN event is tentatively taken into account in \cite{brayeldridge2016, brayeldridge2018, giacobbomapelli2020}.}, thanks to two different possible reasons:

1) some CCSN explosions may occur in an `ultra-stripped' progenitor, already deprived of its hydrogen envelope by previous mass loss and transfer to the companion, so they eject a small mass, often with a low binding energy. In this case, at least the ultra-stripped CCSN with relatively small iron cores may lead to fast explosions and get small kicks; % \cite{tauris2015};

2) a smaller energy of explosion ---because the H--envelope is small---  also occurs in the ecSN due to accretion induced collapse (AIC) of an O--Ne WD \cite{miyaji-nomoto1987},  and in single ecSN, as these latter events occur at the lower mass edge of the SN explosions range. A smaller kick velocity is produced as a consequence. The explosions are under-energetic because there is too little mass to absorb neutrinos ---see e.g. the numerical simulations in \cite{dessart2006}. 

Synthetic computation of NS formation and evolution to MSP in GCs can be found. The recent numerical study  \cite{ye2019}  shows that GCs born with the typical present-day mass of $M_{GC}=2-5\times 10^5$\msun\ may produce up to 10--20\,MSP, while initially more massive GCs of $\sim 10^6$\msun can produce $\sim$100\,MSP. The larger number is to be ascribed both to the larger number of NS formed, keeping the same assumptions on the formation channels, and to the lower escape velocity ($v_e \propto \sqrt{M_{GC}}$). \\
As far as the initial mass of globular clusters is concerned, it is important to point out that a number of recent studies \cite{li-gnedin2019, baumgardt2019, webb2015} have indeed concluded that many clusters must have been initially significantly more massive and lost a significant fraction of their initial mass. Several models for the formation of multiple stellar populations also require globular clusters to be initially more massive (see %Sect.\,\ref{gcmultiple} and 
Gratton et al. \cite{gratton2019} for a review). The MSP production in globular clusters could therefore be significantly larger (see  \cite{ye2019}).

\section{Concepts  of binary evolution}
\label{sec:3}
Interacting binary evolution is a necessary ingredient for the formation of an MSP, and in many cases for the formation of the NS itself.  We summarize here the main properties of binary interaction and take notice of the relevance of some concepts for the case of MSP evolution. \\
%The action of of tides and the possibility of mass transfer between the stars in a binary system was clear already eighty years ago \cite{kuiper1941}, but the first computations could begin much later on. 
The first  important summary review is by Paczynski \cite{paczynski1971ARAA}, whose basic definitions and relevant concepts are still used, also in this short summary. 
%%%%% EARLY FIGURE 2 %%%% fig2
%\begin{figure}[t]
%\includegraphics[scale=0.25]{./figure/hr5crop.pdf}
%\includegraphics[scale=0.25]{./figure/radius5crop.pdf}
%\caption{The classic description of case A, B and C of binary evolution \cite{paczynski1971ARAA} is refreshed by adopting the HR diagram and radius evolution for a 5\msun\ star including the thermal pulse phase of the AGB (enlarged in the inset).  In case A , contact with the Roche lobe occurs in the main sequence phase (1-2), in case B it occurs during the excursion through the Hertzsprung gap (2-3) or during the ascent on the red giant branch (3-4), and in case C it occurs during the AGB evolution (6-7).     }
%\label{fig2}       % Give a unique label
%\end{figure}

\subsection{The Roche lobe and the radius evolution of single stars}
The most important way of how mass can be transferred from one star to the other is the `Roche-lobe overflow' (RLOF). The definition of Roche lobe for a binary component is based on the potential surfaces defined by the gravitational potential of the two orbiting masses plus the centrifugal force acting on a mass-less test particle in a reference frame co-rotating with the binary, in a number of simplifying assumptions  (the orbit has zero eccentricity, the gravitational fields of the stars can be approximated as those of point masses, and stellar rotation is synchronised with the orbital motion). The gravitational potential has 5 `Lagrangian points' where the gradient of the effective potential is zero. %(i.e., where there is no force in the co-rotating frame). 
Three points lie along the line that connects the two stars. The equipotential surface that passes through L1 (the critical Roche-Lobe potential) connects the gravitational spheres of influence of the two stars. If one star begins filling its Roche lobe, matter can
flow through the L1 point into the Roche lobe of the other star. Approximating the Roche lobe volume with a sphere, the Roche lobe radius of the component $M_d$\  ($R_{RLd}$) can be evaluated with the expression by Eggleton \cite{eggleton1983}
\begin{equation}
\frac{R_{RLd}}{a}= { {0.49 q^{-2/3}} \over {0.6q^{-2/3} + ln(1+q^{-1/3})} }
\label{roche}
\end{equation}
where $a$\ is the orbital separation, $q=M_d/M_{acc}$ is the mass ratio, and $M_{acc}$\  is the mass of the companion (the accreting component, if mass transfer takes place).\\
The expression by Kopal  \cite{paczynski1967} is still very useful for $q\simlt 0.5$:
\begin{equation}
\frac{R_{RLd}}{a}= \frac{2}{3^{4/3}}\left( M_d \over {M_d+M_{acc}} \right)^{1/3}
\label{roche-kopal}
\end{equation}
This expression, coupled with Kepler's third law 
\begin{equation}
\left(\frac{2 \pi}{P_{\rm orb}}\right)^2a^3=G(M_d+M_{acc})
\end{equation}
provides a useful relation between the orbital period \Porb, the mass and radius of the donor, in the hypothesis that the donor fills the Roche lobe:
 \begin{equation}
P_{\rm orb}=\frac{9 \pi}{\sqrt{2G}}R_{RLd}^{3/2}M_d^{-1/2} \simeq 9 hr \left(\frac{R_{RLd}}{R_\odot}\right)^{3/2} \left(\frac{M_{d}}{M_\odot}\right)^{-1/2} 
\label{porbmassradius}
\end{equation}
Eq.\,\ref{porbmassradius} is an useful first order evaluation of the donor possible mass-radius relations, depending on its structure (main sequence, giant, white dwarf, mixed H--He degenerate remnant).

%\subsection{The radius evolution of single stars}
A binary in which both component are inside their Roche lobe is `detached'. The life of an interacting binary begins when one of the component becomes a mass `donor', filling its Roche lobe and beginning mass transfer to the companion.
%Figure\,\ref{fig2} shows the Hertzsprung Russell (HR) diagram (left panel) and the radius evolution as a function of age (right panel) for a 5\Msun\ star.  (a low metallicity track is adopted here, courtesy of P. Ventura). This figure defines the 
Three cases of mass exchange are defined by Paczynski \cite{paczynski1971ARAA}. Case A occurs when the donor fills the Roche lobe during the main sequence stage; case B occurs when it is out of the main sequence, or during the red giant branch evolution; case C occurs when the donor is evolving through the AGB that is when it burns hydrogen and helium in a shell
%\footnote{This burning is alternate: hydrogen burns in a stationary way for $\sim$90\% of the time, then Helium ignites abruptly in a non--stationary way (thermal pulses \cite{iben1975}). The radius evolution is complex as shown in the inset of Fig.\,\ref{fig2}.}. 
From Eq.\,\ref{roche} which case  applies depends on the separation $a$\ and on the mass ratio $q$.

\begin{description}
\item{\it Role for MSP:} The historical subdivision in case A, B and C is still important to understand the previous evolution of MSP binaries, whose former donor is now a compact object (see Sect.\,\ref{sec:5}).
\end{description}

%%%%% EARLY FIGURE 3 %%%% rochem
%\begin{figure}[t]
%\sidecaption[t]
%\includegraphics[scale=0.35]{./figure/rochemcrop.pdf}
%\caption{Roche lobe radius in conservative evolution starting from an initial separation $a_{in}$=4.94$R_\odot$\ for two examples with different total mass. A possible donor evolves at constant mass, increasing radius, until it meets the Roche lobe. Thereon, the evolution depends on whether the stellar equilibrium radius $R_{eq}$\ remains inside the lobe (evolution on the nuclear timescale), as from the point N, which can follow the Roche lobe conservative evolution ---red squares), or it becomes larger, and the evolution is on the thermal timescale (T), or, if the adiabatic radius also becomes bigger than the lobe radius, the system is unstable to dynamic mass transfer (D).  }
%\end{center}
%\vskip -70pt
%\label{rochem}       % Give a unique label
%\end{figure}

\subsection{The radius change due to mass loss }
\label{drdm}
The most important property of mass transfer is its stability, and this depends on the comparison of the response of the stellar radius, to the concomitant response of the Roche lobe radius  to mass loss. We define the three {\sl mass-radius exponents}, from which three different {\sl timescales of mass transfer} depend.
%reaction of both donor and accretor to mass loss, mainly determined on whether the stellar envelope is radiative or convective. 
%%%%% FIGURE 3 %%%% rochem
\begin{figure}[t]
\sidecaption[t]
%\vskip -30pt
%\begin{center}
\includegraphics[scale=0.35]{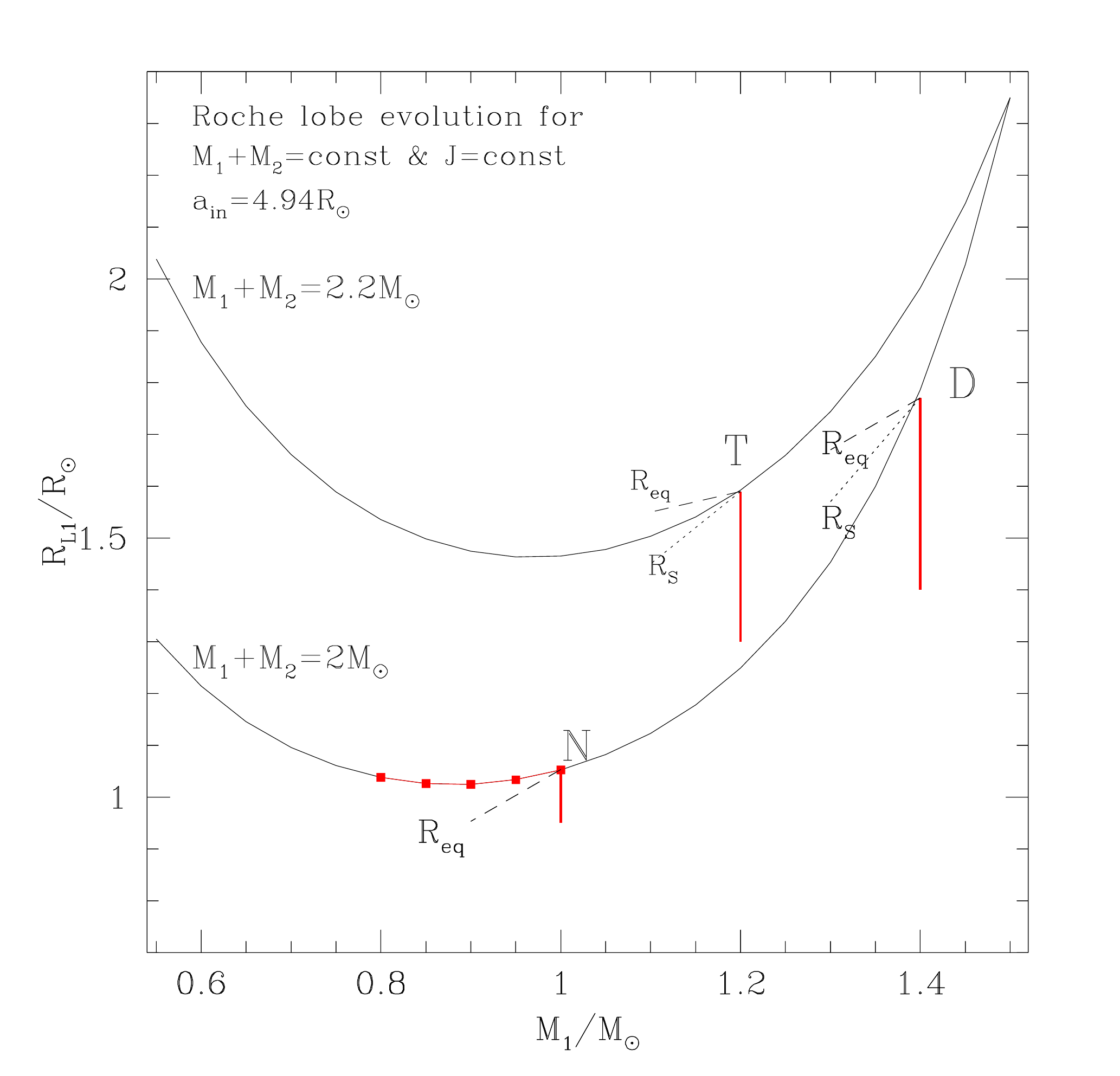}
%\vskip -10pt
\caption{Roche lobe radius in conservative evolution starting from an initial separation $a_{in}$=4.94$R_\odot$\ for two examples with different total mass. A possible donor evolves at constant mass, increasing radius, until it meets the Roche lobe. Thereon, the evolution depends on whether the stellar equilibrium radius $R_{eq}$\ remains inside the lobe (evolution on the nuclear timescale), as from the point N, which can follow the Roche lobe conservative evolution ---red squares), or it becomes larger, and the evolution is on the thermal timescale (T), or, if the adiabatic radius also becomes bigger than the lobe radius, the system is unstable to dynamic mass transfer (D). }
%\end{center}
%\vskip -70pt
\label{rochem}       % Give a unique label
\end{figure}

\begin{trailer}{Mass--radius exponents}
%\subsubsection{Mass--radius exponents}
%\begin{enumerate}
%\item  
\noindent
{\sl adiabatic}: $\zeta_S=\left( {\frac{\partial \log R_d} {\partial \log M_d}}  \right)_S$ -- change at constant entropy \\
% (hydrostatic but not thermal equilibrium
%\item  
{\sl equilibrium}: $\zeta_{eq}=\left( {\frac{\partial \log R_d} {\partial \log M_d}}  \right)$ -- change preserving hydrostatic and thermal equilibrium
%\item  
\noindent
{\sl Roche lobe}: $\zeta_{RLd}=\left( {\frac{\partial \log R_{RLd}} {\partial \log M_d}}  \right)$
%\end{enumerate}
\end{trailer}
%Then the mass transfer (\cite{webbink1985}) may occur on three different timescales: %according to its occurrence on the nuclear, thermal or dynamical timescale. 
\begin{trailer}{Timescales of mass transfer}
\noindent
{\it \underline{nuclear}:} $\zeta_{RLd} < min( \zeta_S, \zeta_{eq}$). Mass transfer is stable, mass loss is due to the {\it expansion} of the donor caused by nuclear evolution,  while the donor remains in hydrostatic and thermal equilibrium. The approximate timescale (for core--H burning) is 
%\begin{equation}
$\tau_{nuc} \sim 10~{\rm Gyr} \frac{M}{M_\odot}\frac{L_\odot}{L} $.
%\end{equation} 
If $\tau_{nuc}$\ is too long (for masses having a timescale of hydrogen burning longer than the age of the  Universe), systemic angular momentum losses may act as a driver, on the timescale of {\it contraction} of the Roche lobe radius; %(see Eq.\,\ref{tauaml});
\\
\noindent
{\it \underline{thermal}:}   ~~$\zeta_S>\zeta_{RLd} > \zeta_{eq}$~~ Mass transfer is driven by the expansion of the donor, for example when it crosses the Hertzsprung gap after the end of the core--hydrogen burning phase. The star remains in hydrostatic equilibrium, and the mass loss is limited by the thermal readjustment at a maximum value $\sim \dot{M}_{max}= - M_d/ \tau_{KH}$, where %An approximate expression is % $\tau_{KH}$\  
%\begin{equation}
 %\tau_{KH} \sim \frac{GM^2}{2RL} \sim 15~{\rm Myr} \left( \frac{M}{M_\odot} \right)^2 \frac{R_\odot}{R}  \frac{L_\odot}{L} 
%\label{tkh}
%\end{equation} 
 $\tau_{KH} \sim (GM^2)/(2RL) \sim 15~{\rm Myr}( M/M_\odot)^2 (R/{R_\odot}) (L_\odot/L) $
%\item

\noindent
{\it \underline{dynamical}:}   ~~$\zeta_{RLd} > \zeta_S$~~ The adiabatic response of the donor is unable to keep it within the Roche lobe. The growing mass transfer rates reach values $\dot{M}_{max}= - M_d/ \tau_{dyn}$. %, where $\tau_{dyn}\sim 5\times 10^{-5} \sqrt{(R^3/M^3)}$, M and R in solar units. 
The evolution is catastrophic, as the star reacts to perturbations to hydrostatic equilibrium on the dynamical timescale 
%$\tau_{dyn}$. An approximate evaluation is 
%\begin{equation}
%\tau_{dyn} \sim \sqrt{R^3/GM} \sim 0.04\left(\frac{M_\odot}{M} \right)^{1/2} \left(\frac{R_\odot}{R} \right)^{3/2} yr. $
%\end{equation} 
%Fig.\,\ref{rochem} exemplifies these three cases.\\
$\tau_{dyn} \sim \sqrt{R^3/GM} \sim 0.04({M_\odot}/{M})^{1/2} ({R_\odot}/{R})^{3/2} yr. $

Fig.\,\ref{rochem} exemplifies these three cases\footnote{inspired by Figure 7.1 in the lectures on binary evolution by  Onno Pols, at \\
http://www.astro.ru.nl/~onnop/education/}. 
%%%ASTROPH--- POLS
\\
\end{trailer}

\noindent
A typical situation in which dynamical mass transfer ensues is the case of an evolving giant beginning to transfer mass to a less massive companion, as $\zeta_S\sim -1/3$\ for the convective giant envelope, while $\zeta_{RL}>0$\ (the system becomes tighter). This leads to the ``common envelope" (CE) evolution \cite{paczynski1976ce}
% whose final outcome is generally estimated with simple approximations.
%The common envelope 
that is assumed to be responsible, e.g., for the formation of cataclysmic binaries \cite{paczynski1976ce}, and of the double WD merging leading to Type I supernovae \cite{webbink1984, ibentutukov1984}.
Unfortunately, quantitative computations have been rare \cite{meyer1979}, as the spiraling-in process of two stars (or a star and a core) embedded in a CE involves a large number of hydrodynamic and themodynamic processes, occurring on a wide range of time and length scales. Simple expressions have been worked out and widely adopted in the literature. The common assumption is that the gravitational energy lost from the orbit when the mass losing star spirals in towards the companion  is partially deposited into rotation and partially into heating of the envelope, and, in principle, it can cause the envelope ejection when it exceeds the binding energy. Thus an efficiency parameter  is defined as:
%\begin{equation}
$\alpha_{ce} = \frac{\Delta E_{\rm binding}}{\Delta E_{\rm orb}}$
%\end{equation}
and in the simplest case it is evaluated by:
\begin{equation}
\alpha_{ce} \frac{GM_cM_2}{a_f}= \frac{GM^2_{1}}{a}
\end{equation}
where $M_1$\ and $M_2$\ are the initial masses of donor and companion, $M_c$ is the core mass of the donor (for example the mass of the helium core in case B Roche lobe overflow) and $a$\ and $a_f$\ are the initial and final separations before and after the CE \cite{ibentutukov1984}. More elaborated expressions are available (e.g. \cite{liviosoker1988}), but the parametrization remains highly uncertain. \\ 
%\begin{description}
%\item{\it Role for MSP:} CE evolution  is also met when we trace the binary evolution leading to the formation of millisecond pulsars from intermediate and high mass binaries (Sect.\,\ref{DNS}  and \ref{IMtoMSP}). Two CE phases are needed to explain the double NS binaries, so we must be aware that these evolutionary paths are often affected by an uncomplete parametric description.
%\end{description}

%\subsection{The total radius variation}
%We have defined  the partial derivatives with respect to the mass loss. 
The total derivative of radius and Roche lobe radius with respect to time can be written as:
\begin{equation}
\frac{d \ln R_d}{dt}= \zeta_S \frac{d \ln M_d}{dt}+\left( \frac{\partial \ln R_d}{\partial t} \right)_{\dot{M}=0}
\label{der1}
\end{equation}
\begin{equation}
\frac{d \ln R_{RLd}}{dt}= \zeta_{RLd} \frac{d \ln M_d}{dt}+\left( \frac{\partial \ln R_{RLd}}{\partial t} \right)_{\dot{M}=0}
\label{der2}
\end{equation}
The radius varies {\it independently} of mass loss on its timescale 
%\begin{equation}
$\tau_{R_d}=(\partial \ln R_d/ \partial t)_{\dot{M}=0}^{-1}$
%\label{taurelax}
%\end{equation}
due both to thermal relaxation (a contraction, or an expansion) and nuclear evolution (generally an expansion). If the system is isolated, but  primary angular momentum ($J$) loss mechanisms are present (for example, magnetic braking and gravitational radiation), also the Roche lobe radius varies in time independently of mass loss, on the timescale 
%\begin{equation}
$\tau_J={J}/{\dot{J}=2(\partial \ln R_{RLd}/ \partial t)}^{-1}_{\dot{M}=0}$
%\label{tauaml}
%\end{equation}

%\begin{description}
%\item{\it Role for MSP:} We will see 
%In Sect.\,\ref{radvarillum} we will have to consider another important radius variation, acting on the thermal timescale of the donor convective envelope, which alters significantly the evolution in the case of binaries containing an MSP.
%\end{description}

\subsection{The losses of angular momentum: primary mechanisms }
\label{amloss}
The orbital angular momentum of binaries $J$\ can be written as:
\begin{equation}
J^2= J^2_{\rm orb}=G \frac{M_d^2M_{acc}^2}{M} a (1-e^2)
\end{equation}
where $G$\ is the gravitational constant, $e$\ the system eccentricity and the total mass is $M=M_d+M_{acc}$. Differentiating we obtain the general orbital evolution equation
\begin{equation}
2\frac{\dot{J}}{J}= \frac{\dot{a}}{a} + 2 \frac{ \dot{M_d}}{M_d} + 2 \frac{ \dot{M}_{acc}}{M_{acc}} - \frac{\dot{M}}{M}  - \frac{2e \dot{e}}{(1-e^2)}
\label{amder}
\end{equation}
For $e=0$\ and conservative mass transfer ($\dot{J}=0$\ and $\dot{M}_{acc}=-\dot{M_d}$), Eq.\,\ref{amder} reduces to:
\begin{equation}
\frac{\dot{a}}{a} = - 2 \left( 1- \frac{M_d}{M_{acc}} \right) \frac{ \dot{M_d}}{M_d} 
\label{amdercons}
\end{equation}
As $\dot{M_d} <0$, we find the well known result that the separation {\it increases} when the donor is the less massive component, and {\it decreases} in the opposite case. A simple way to understand this result is to see it from the point of view that total AM is constant: mass transfer to the more massive component brings matter {\it closer} to the center of mass, to a status of smaller AM, thus  AM must be added to the orbit, which becomes wider. The reverse occurs when matter is transferred to the lighter component, and thus has to achieve AM from the orbit. 

In the general case, we must consider all the possible angular momentum variations. We can write
\begin{equation}
\frac{\dot{J}}{J}= \frac{\dot{J}_{GR}}{J} + \frac{\dot{J}_{MB}}{J} + \frac{\dot{J}_{ml}}{J}
\end{equation}
The two first terms are primary losses, as they act also when the binary is detached, the third one is the consequence of mass loss. \\
%\begin{description}

\noindent
{\bf Magnetic braking: (MB AML)} ${\dot{J}_{MB}}/{J}$. The single stars having a convective envelope ($M \lesssim 1.3M_\odot$) are known to slow down during their main sequence lifetime \cite{skumanich1972, soderblom1983}. The presence of the convective envelope produces a dynamo magnetic field, and turbulence and wind mass loss.  Schatzman \cite{schatzman1962} pointed out that  the winds of these stars are magnetically constrained to corotate with the star, out to distances large compared with the stellar radius, so that even a very small amount of mass loss would yield a proportionally much greater loss of angular momentum. This `magnetic braking' is schematized in various parametric ways in the literature, mainly calibrated on the Skumanich relation \cite{skumanich1972} that the stellar angular velocity decreases as the age at power -1/2. In \cite{vz1981} the expression (in cgs units) is:
\begin{equation}
 \frac{\dot{J}_{MB}}{J} \simeq -0.5 \times 10^{-28} f^{-2}_{mb} \frac{k^2R_d^4}{a^5} \frac{GM_d^3}{M_dM_{acc}}  s^{-1}
\label{JMB} 
\end{equation}
where $R_d$\ is the radius of the mass-losing star; $k^2$\ is its gyration radius and $f_{mb}$\ is a constant of order unity. Several other (parametric anyway) formulations have been tested and adopted in the literature. \\

%\item
\noindent
{\bf Gravitational radiation (GR AML)} ${\dot{J}_{GR}}/{J}$. From general relativity
\begin{equation}
 \frac{\dot{J}_{GR}}{J} = \frac{32 G^3}{5c^5} \frac{M_dM_{acc}M}{a^4} s^{-1}
\label{JGR} 
 \end{equation}
where c is the speed of light in vacuum. This term is especially significant for close orbits, and in particular for short period cataclysmic binaries.
%\end{description}
\noindent
%The systemic MB and GR AML  are the driving mechanisms of evolution 
%in CBs, in which the accreting WD component of mass $M_{WD}$\ is subject to several  explosive phenomena, while the donor star is in most cases a quasi--main sequence low mass dwarf having  $M_d < M_{WD}$.
%The nuclear timescale of evolution of these objects is well beyond the age of the Universe, so their radius increase can not be the reason for losing mass. 
%Even if the system is in contact at some stage, according to Eq.\,\ref{amdercons} mass transfer at constant angular momentum would increase the separation and  the dwarf would find itself again inside the Roche lobe.
%Thus the evolution of CBs needs {\it primary} angular momentum losses (AML) from the binary.
%Note that the emission of 
Note that gravitational radiation is a physical consequence of the general relativity, but the magnetic braking 
%is an empirical, although well sound, mechanism, so its 
formulation resides on parameters and functional dependences which are, at best, semiempirical. In fact, calibration of the magnetic braking losses is done by comparison of the results of binary evolution with the properties of the relevant classes of objects. \\

%\subsection{Angular momentum losses linked to the mass transfer: `consequential' AML (CAML)}
\noindent
{\bf Consequential AML} $ {\dot{J}_{ml}}/{J}$: 
If the mass transfer is non conservative, and a fraction $\alpha$\ of mass is lost by the system, this mass will carry away angular momentum, and we can write:
\begin{equation}
% \frac{\dot{J}_{ml}}{J} = \alpha \dot{M} j_{spec}
 \dot{J}_{ml} = \alpha \dot{M}  j_{spec}
\end{equation}
where $j_{spec}$\ is the specific AM of the mass leaving the system. It is useful to write again $J_{orb}$, for circular orbit in terms of the orbital angular velocity $\omega=\sqrt{(GM/a^3)}$:
\begin{equation}
J_{orb}=\frac{M_d M_{acc}}{M}a^2 \omega
\end{equation}
to explicit how $j_{spec}$\ is written in terms of the total AM of the binary.
A typical case considered in the literature is the loss of AM occurring when a nova explosion expels the accumulated hydrogen rich envelope. In this case we can make the assumption that the nova shell is lost with the specific angular momentum of the WD \cite{schenker1998}. In cases of mass transfer rates larger than the Eddington limit, a typical assumption is that the mass lost by the donor ($\dot{M}$) is acquired by the accretor up to the Eddington mass accretion rate ($\dot{M}_{Edd}$ in modulus) but the rest is lost with the specific AM of the accretor $j_{acc}= a^2_{acc}\omega=[M_d /(M_{acc}M)]J_{orb}$ (being $a_{acc}=(M_d/M)a$\  the distance of $M_{acc}$\ from the center of mass), and:
\begin{equation}
 \frac{\dot{J}_{ml}}{J} = \frac{M_d}{M_{acc}M}  (\dot{M}+|\dot{M}_{Edd}|) 
\end{equation}
In the case of wind mass loss from the donor, the associated specific AML is probably that of the donor, while, specifically for low mass companions of MSP in the radio-ejection phase, it has been proposed that the mass lost by this mechanism carries away the specific AM at the inner lagrangian point L1, which is  $j_{L1}=d^2_{L1}\omega$, where $d_{L1}$\ is the distance between L1 and the binary center of mass.  In this case we can write:
\begin{equation}
 \frac{\dot{J}_{ml}}{J} = \frac{M}{M_d M_{acc}} \frac{d^2_{L1}}{a^2} \times \alpha \dot{M} 
\end{equation}
where the fraction is $\alpha$=1 for radio--ejection. 
%following Alessandro
Note that, fixed $\dot{M}$, the AM lost  is minimum if the mass is lost from the accretor (which is the mass closer to the barycentre) much larger if the loss is from L1, and a bit larger if the loss is directly from the donor.

\begin{description}
\item{\it Role for MSP:} MB and GR AML are both considered in the secular binary evolution of short period binary MSP. A role for consequential AML is probably played both during the super--Eddington phases of mass transfer and during the mass loss induced on the companion by the action of the MSP radiation, both in the case of evaporation and in the case of radio-ejection. Whatever the choice for the specific AML associated to mass loss from the system, the  term is  so important that it may qualitatively affect the result of binary evolution.

\end{description}

\subsection{The approach of a donor to the Roche lobe contact}
\label{mdotdot}
Stars have extended atmospheres, so mass flow from the donor begins when the photospheric radius is still smaller than the Roche lobe radius. While the mass flow through the inner Lagrangian point is a complicated hydrodynamical problem, a simple expression to estimate it in the `optically thin' case was developed by Ritter \cite{ritter1988}.
\begin{equation}
\dot{M} = - \dot{M_0} \exp\left({(R_d- R_{RLd} \over{H_p}}\right)
\label{massloss}
\end{equation}
and
%\begin{equation}
$ \dot{M_0}={ (1/\sqrt e)}\rho_{ph} v_s Q$,
%\end{equation}
where $v_s$\ and $Q$\ are the isothermal sound speed and the effective cross section of the flow at L$_1$, and $\rho_{ph}$\ is the photospheric density.\\
\noindent
Eq.\,\ref{massloss} shows that the mass transfer rate increases exponentially while the stellar radius approaches the Roche lobe, until the stationary situation is reached, when $R_d=R_{RLd}$ and  $dR_d/dt=dR_{RLd}/dt$. In \cite{ritter1988} we find the discussion of the evolution of the mass transfer rate and in \cite{dantona1989} numerical examples of the different approach to stationary mass loss for different initial donor masses. \\
We can 
%recover the equation \ref{stationarymdot} for the stationary mass transfer rate by 
explicitly write down the acceleration by differentiating Eq,\,\ref{massloss}, and assuming that $\dot{M}_0$\ is constant:

\begin{equation}
\ddot{M}= \dot{M} \frac{R_d}{H_p} \left[ \frac{d \ln R_d}{dt} - \frac{d \ln R_{RLd}}{dt} \right]
\label{accmloss}
\end{equation}
Substituting the derivatives as written in Eq.\,\ref{der1} and \ref{der2}, we write:
\begin{equation}
\ddot{M}= \dot{M} \frac{R_d}{H_p} \left[  \left( \zeta_S - \zeta_{RLd} \right) \frac{d \ln M_d}{dt} +
\left( \frac{\partial \ln R_d}{\partial t} \right)_{\dot{M}=0} + \left( \frac{\partial \ln R_{RLd}}{\partial t} \right)_{\dot{M}=0} \right] 
\label{ddotmloss}
\end{equation}
 and using the definition of thermal relaxation and AML timescales, %(Eq.\,\ref{taurelax} and \ref{tauaml}), 
% \begin{equation}
%\ddot{M}= \dot{M} \frac{R_d}{H_p} \left[  \left( \zeta_S - \zeta_{RLd} \right) \frac{d \ln M_d}{dt} + \frac{1}{\tau_{R_d}}
%- \frac{2}{\tau_J} \right] 
%\end{equation}
%and, for $\ddot{M}=0$, we go back to Eq.\,\ref{stationarymdot}.
%The timescale of mass transfer is
we get the timescale of mass transfer:
\begin{equation}
\tau_{\dot{M}} = \frac{\dot{M}}{\ddot{M}} = \frac{H_p}{R_d} \left[ {\frac{\dot{M}}{M_d}}(\zeta_S-\zeta_{RLd})+ {\frac{1}{\tau_{R_d}}} - {\frac{2}{\tau_J}}  \right] ^{-1}
\label{timescaleddotm}
\end{equation}
At first, when the donor is still well within its Roche lobe, the dominant term in Eq.\,\ref{timescaleddotm} is the third one, the timescale of angular momentum loss, and
\begin{equation}
\tau_{\dot{M}} = \frac{\dot{M}}{\ddot{M}} \sim \frac{H_p}{2 R_d} \tau_J
\label{tauin}
\end{equation}
As soon as the radius reaction to mass loss becomes effective, the first term enters into play, as well as the thermal relaxation term. In the stable case, at the end the two first terms balance the third one and a stationary value for \Mdot\ is achieved %and we recover Eq.\,\ref{stationarymdot}. \\

\begin{equation}
-\dot{M_d}= {\frac{M_d}{\zeta_S-\zeta_{RLd}}} \left[ {\frac{1}{\tau_{Rd}}} - {\frac{2}{\tau_J}} \right]
\label{stationarymdot}
\end{equation}
The stationary mass transfer in the conservative case is also  found directly from Eq.\,\ref{der1} and \ref{der2}, by requiring that the radius of the donor and of  the Roche lobe, and their respective time derivatives are equal. 
Notice that Eq.\,\ref{stationarymdot} requires that $\zeta_S > \zeta_{RLd}$ for dynamically stable mass transfer (Sect.\,\ref{drdm}).\\
%There are many cases in which the onset of mass transfer is a phase which may significantly alter the successive evolution, so it must be explicitly considered.\\

The approach to the stationary rate thus depends on the characteristic timescales of the donor under consideration. 
%An exemplification of the differences as a function of the donor mass is discussed in \cite{dantona1989}. 
%In Sect.\,\ref{radvarillum} we will see how the onset and the secular evolution is modified by additional terms entering the play. Notice that 
The timescale of non--stationary mass transfer (Eq.\,\ref{tauin}) is only $H_p/2 R_d \simeq 10^{-4}$ of the secular evolution timescale $\tau_J$.
\begin{description}
\item{\it Role for accreting MSP}: primary losses of AM dominate the timescale and average mass transfer rate for binaries with scarcely evolved companions. If, for any reason, the mass transfer timescale becomes shorter (and thus \Mdot\ larger) a  lower \Mdot\ at a following epoch will be  necessary. In CBs the irradiation will produce short duration `limit cycles' (King et al. \cite{king1995cycles}). In LMXBs this produces cycles of mass transfer followed by fully detached stages (see Sect.\,\ref{xraycycles}). 
\end{description}

\section{Cataclysmic binaries as a comparison key study}
\label{sec:4}
%The interacting binaries containing a WD accretor include a variety of objects, depending of the nature of the donor companion, %\footnote{In the AM CVn the donor is another WD; in the symbiotic stars the donor is a red giant},
%but the most populated class is that of Cataclysmic Binaries (CBs), where the donor is a low mass ($\lesssim$\,1\msun) dwarf. 
We summarize here the concepts framing the secular evolution of Cataclysmic binaries (CBs), semi--detached systems consisting of a quasi--main--sequence donor and a WD companion, because they 
%has been studied at long, to understand their optical variability events, easily monitored in the sky, produced by accretion  onto the WD.  It is worth to sketch the main scheme of evolution deviced for CBs, because it has been 
can be (in part) applied to study the evolution of binaries with NS companions.\\
%%%%% EARLY FIGURE 4 %%%% ritter
%%%%% FIGURE 2 %%%% ritter
\begin{figure}[t]
%
%\sidecaption[t]
\begin{center}
\vskip -45pt
\includegraphics[scale=0.5]{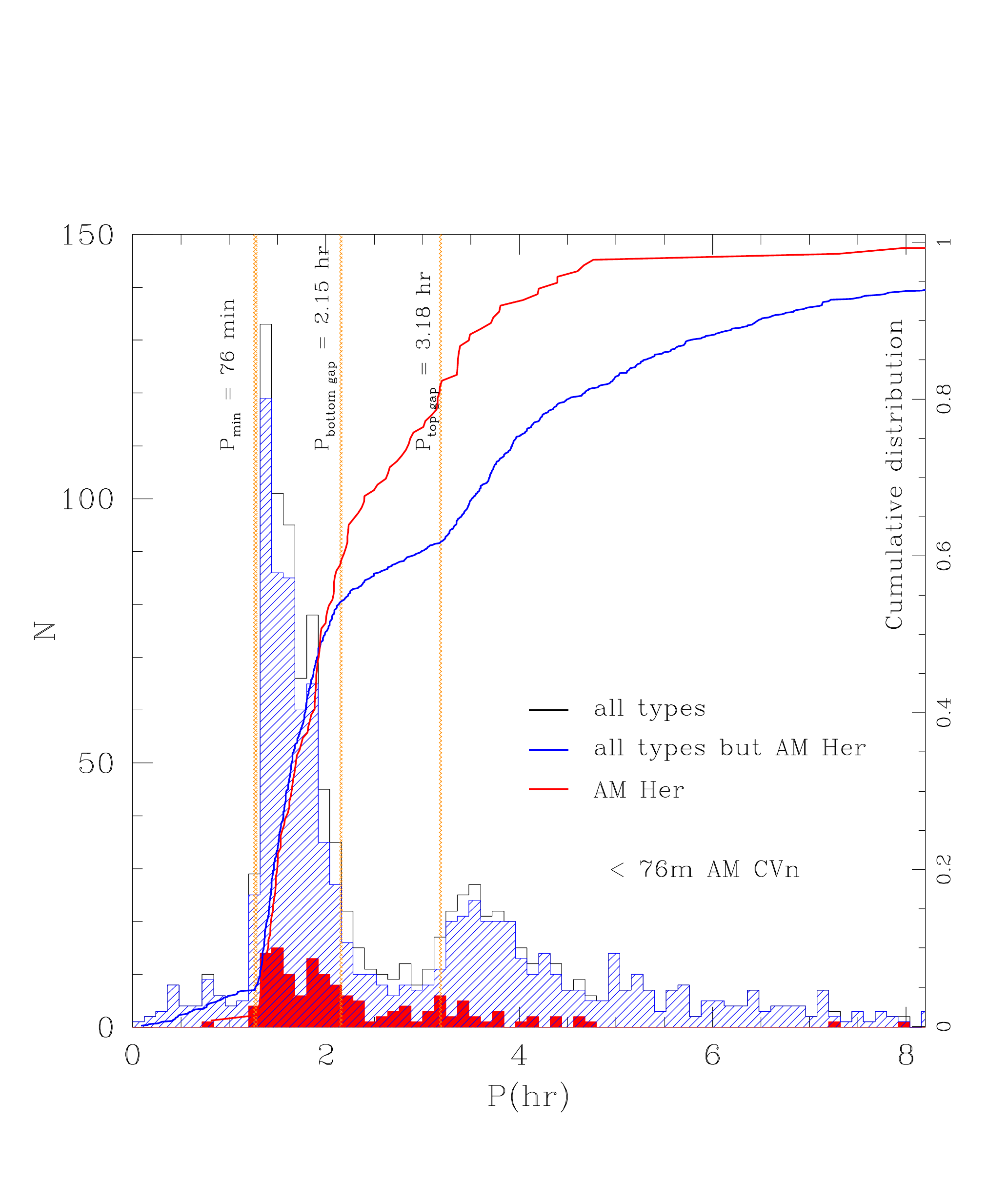}
\vskip -20pt
\end{center}
\caption{ Period distribution of CBs from the 7th Edition, Release 7.21 (March 2014) of the Catalog by Ritter \& Kolb \cite{ritterkolb2003}. The cumulative distribution is normalized to the total numbers in the catalogue, but only periods up to 8\,hr are shown. The differential distribution shows separately the Polars (AM\,Her) and the sample not including the polars. The minimum period and the boundaries of the 2-3\,hr period gap are highlighted. The gap is well defined by the flattening in the cumulative distribution of CBs, but is not present in the distribution of Polars. 
%The plot has been inspired by a similar display by Knigge \cite{knigge2011rev}.
} 
%\vskip -60pt
\label{ritter}       % Give a unique label
\end{figure}
%\subsection{The standard evolution of cataclysmic binaries}
At \Porb$\lesssim$10\,hr the donor of CBs must be a scarcely evolved quasi--MS star, thus  primary AML is required to keep the Roche lobe radius in constant contact with the donor radius, allowing stable mass transfer. 
A `standard' model has been developed to account for the observational properties of different classes of CBs. In particular, the non-magnetic CBs shows a shortage of stars at orbital periods between $\sim$3.2\,hr and $\sim$2.15\,hr (period gap), as shown in Fig.\,\ref{ritter}. \\
%Alessandro follows
We can write the mass--radius relation as a power law:
\begin{equation}
\frac{R}{R_\odot} = f \left(\frac{M_d}{M_\odot}\right)^\alpha
\label{massrad}
\end{equation}
where $f \sim 1$\ and $\alpha \sim 1$\ for low mass main sequence stars. We can get rid of the dependence of the orbital period on the radius of the mass-losing component  in Eq.\,\ref{porbmassradius}, and obtain:
\begin{equation}
P_{\rm orb}^2 \propto M_d^{3\alpha-1}
\end{equation}
The logarithm derivative provides:
\begin{equation}
\frac{\dot{P}_{\rm orb}}{P_{\rm orb}} = \frac{3\alpha-1}{2} \frac{\dot{M_d}}{M_d}
\label{pdotmin}
\end{equation}

The typical average mass transfer rates observed above the gap are $\sim 10^{-9}$\msun/yr, too large to be provided by GR AML, while they can be met with a reasonable choice of parameters of MB AML from Eq.\,\ref{JMB}. For a typical mass of 0.4\msun\ the mass loss timescale is $\sim 0.4/10^{-9} \sim 5\times 10^8 yr$, similar to the thermal timescale. %(from Eq.\,\ref{tkh}). 
Such a mass loss keeps the donor out of thermal equilibrium, meaning that its radius remains somewhat larger than the MS radius (or that $\alpha <1$\ in Eq.\,\ref{massrad}). The  best model to describe the presence of the period gap is to recognize that the donor becomes fully convective at \Porb$\sim 3$\,hr\footnote{At \Porb=3\,hr the donor has a mass $\sim$0.2\Msun. In MS models, full convection is reached at $\sim0.35$\Msun, but thermal disequilibrium modifies the structure and these mass losing models achieve full convection at a smaller mass.}. This change of structure corresponds to a readjustment of the magnetic field in a way that  MB AML becomes suddenly much smaller, and allows the donor to contract back to thermal equilibrium. Thus the donor detaches from the Roche lobe and mass transfer stops. It is mainly GR AML that, at this stage, will shrink the orbit and bring the donor again into contact at a period $\sim$2\,hr. The ``disrupted magnetic braking" model was first proposed by Ritter \cite{ritter1985}.
AM Her or polar CBs (in which the spin of a highly magnetized WD is locked with the \Porb) are found  in the period gap in the proportion expected if there is no discontinuous behaviour. MB is probably less efficient in these systems, because synchronous rotation allows the connection between the lines of the magnetic fields of donor and of the WD, and there are few open lines left for the wind to carry away angular momentum \cite{webbinkwick2002}. \\
Below the gap, the evolution of CBs resumes and proceeds at lower mass loss rate, \Mdot$\sim$few$\times 10^{-11} $ \Msun/yr. This can be easily provided by pure GR AML. Even the timescale for such a low mass transfer rate becomes shorter than the thermal timescale, and the low mass donor is driven out of thermal equilibrium and preserves a larger radius.  Electron degeneracy becomes increasingly important in the structure approaching the transition to brown dwarf, the mass radius exponent $\alpha$\ lowers, the period reaches a minimum when $\alpha \sim 1/3$\ (Eq.\,\ref{pdotmin}), and begins increasing again. Nuclear burning is still active at the minimum period, even if the stellar mass is well below the `standard' minimum mass for core hydrogen burning ($M \sim 0.08$\msun), because the structure is not in thermal equlibrium and thus keeps a larger central temperature. Paczynski \cite{paczynski1981}, \cite{paczynski-sienkiewicz1981} proposed that the minimum period observed in the period distribution of CBs at 80\,min (actually at $\sim$76\,min, according to the more complete modern inventory, see Fig.\,\ref{ritter}) is due to GR AML.
%According to recent  modelling,  a residual MB AML may still be necessary to achieve a good quantitative agreement with the observed \Pmin \cite{knigge2011}. \\
Thus, in spite of the need to calibrate the parameters in the description, the qualitative (and quantitative) picture of the secular evolution of CBs is settled. For a fresh and updated summary see \cite{knigge2011rev}. 

\subsection{Extension of the CB evolution scheme to X--ray binaries}
\label{binevol}
While the scenario for the evolution of CBs was building up, some bright  galactic X-ray sources were discovered in binaries in the hours period range, and appeared to be the counterparts of CBs, where a NS accretor had replaced the WD\footnote{Interesting to note that two of the few X-ray binaries with determined periods hinting for the analogy with CB evolution were 4U 1626--67 \cite{middleditch1981}, with \Porb=41\,min, well below \Pmin, and Cyg\,X3, with a \Porb=4.8\,hr. This latter, in spite of its `typical' period of a few hours, is indeed the final stage of a high mass interacting binary, a black hole accreting from a mass losing Wolf-Rayet star \cite{vdhdeloore1973}. }. It was then a natural extension to adopt a similar secular evolution scheme \cite{rjw1982}. 
%\cite{rjw1982, podsirappaportpfhal2002}. 
%An interesting feature noted in \cite{podsirappaportpfhal2002} is that, if the parameters of evolution  for MB AML are kept at the values necessary to reproduce the period gap of CBs, a tighter period gap results when the WD is replaced with a NS companion. This is due to the  factor $\sim$2 larger mass of an NS with respect to the average WD mass, so the AML will be a half (see Eq.\,\ref{JMB}), and the donor will be more massive (with a lower thermal disequilibrium) when it becomes fully convective. Obviously the parameters in the magnetic braking law can be different in these two classes of binaries. 
%
%\subsubsection{The bifurcation period }
%\label{bifurcation}

In this context, Tutukov et al. \cite{tutukov1985} gave the first description of the `bifurcation' period (\Pbif) between systems evolving towards short or long \Porb. The evolution arrow depends on which timescale prevails. If  
the AML prevails on the H--core nuclear burning ($\tau_J < \tau_{nuc}$) mass loss will erode the non evolved MS star, in the typical secular evolution described above; if the H--core burning prevails ($\tau_J > \tau_{nuc}$) mass loss can not prevent the formation of a H--exhausted core, the stellar radius increases and the system follows a standard `case B' evolution.
%Depending on the initial mass ratio, \Porb\ may initially decrease due to mass loss occurring on the thermal timescale until $M_d > M_{acc}$\, but the evolution will continue on the nuclear timescale as soon as the mass ratio is reversed, \Porb\ increases until the star has such a tiny H--envelope left that it contracts within the Roche lobe. 
%In \cite{pylyser1988, ergma1996,  taurissavonije1999} 
In \cite{pylyser1988, ergma1996}  we find the first discussions of binary evolution above \Pbif\ for the MSP case, and see Sect.\,\ref{scarcely}.

%\subsubsection{Ultra-short period binaries}
%\label{ultrashort}
Systems starting mass loss at periods close to the bifurcation have the arrow of evolution pointing towards shorter periods, but an advanced core hydrogen exhaustion, or even a very small helium core ($\sim 0.02-0.03$\msun) is already present. The radius for a fully degenerate hydrogen--helium core with hydrogen mass fraction $X$\ can be approximated (\cite{paczynski1967}) by:
\begin{equation}
\frac{R_d}{R_\odot} =0.013 (1+X)^{3/2} \left(\frac{M_d}{M_\odot}\right)^{-1/3}
\end{equation}
%(a better analytic approximation can be found in \cite{nelson-rappaport2003}). 
Thus the stellar radius is smaller for each given mass. According to Eq.\,\ref{porbmassradius}, these systems may evolve to `ultrashort' periods \cite{fedorovaergma1989}.
%\cite{fedorovaergma1989, podsirappaportpfhal2002}. 
The  ultracompact LMXBs 4U\,1626--67 (\Porb=41\,min \cite{middleditch1981}) and 4U\,1915--05  (\Porb=50\,min \cite{chou2001}) should have degenerate donors remnants of cores in which hydrogen has been partially depleted. 
%The shortest period LMXB 4U\,1820--30 in the globular cluster NGC 6624 \cite{stella1987} is better explained as result of a tidal capture of a MS star, followed by a CE phase \cite{bailyngrindlay1987}, but the evolution described here may be an alternative scenario. 
The formation of ultracompact binaries in the field may also be due to different pathways through intermediate mass X-ray binaries evolution for a He-star plus NS \cite{chen2016}.
%The population of close binary WDs below \Pmin\ (see Fig.\,\ref{ritter}) is made up mostly by AM\,CVn binaries, in which a very low mass helium WD transfers matter to the more massive WD accretor. The formation of these objects too may have followed different paths, including the case of primaries starting mass loss close to, but below, the bifurcation period. 
%(see also \cite{pylyser1988,  ergmasarna1996}). 

\subsection{Comparisons and the birthrate problem}
\label{comparison-classes}
\begin{figure}[t]
%\sidecaption[t]
\vskip -30pt
\center{
\includegraphics[scale=0.26]{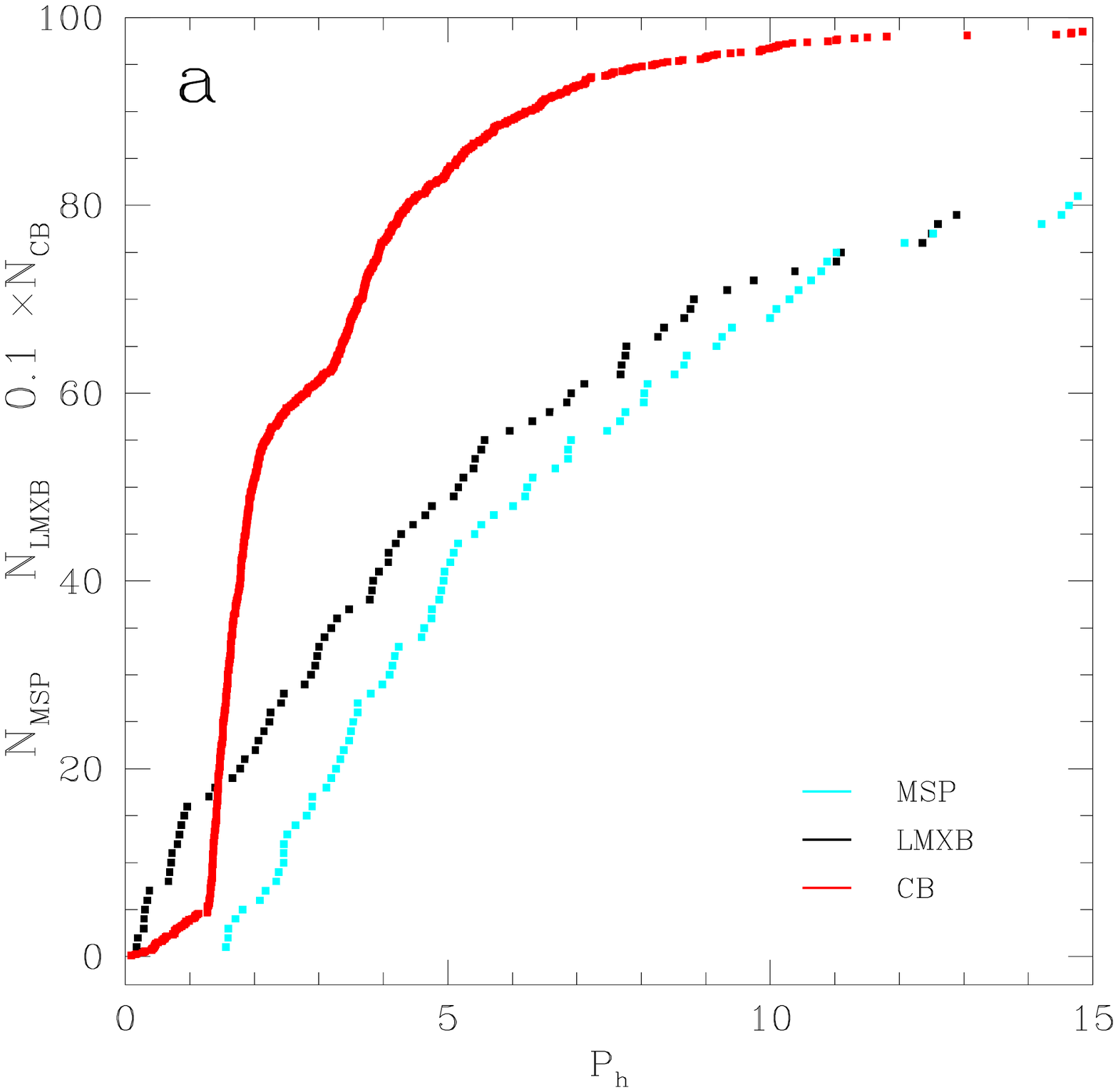}
\includegraphics[scale=0.26]{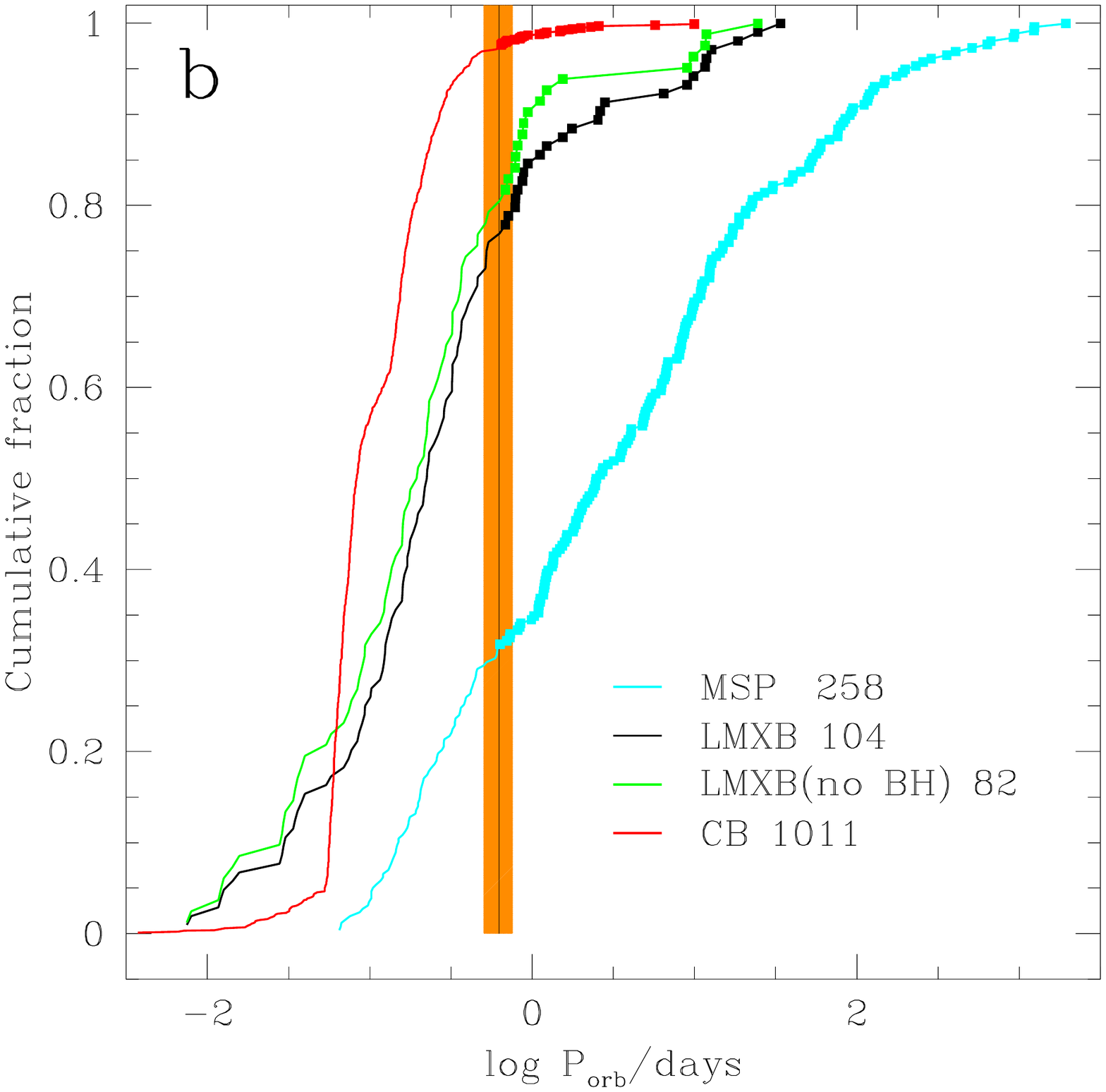}}
\vskip -40pt
\caption{Panel a: comparison of the cumulative number counts for LMXB, MSP and CBs as a function of \Porb in the range 0--15\,hr. Note the different scale for the very abundant CB sample.
Panel b: the same comparison highlights with dots the systems above the bifurcation period (orange band).  Only periods up to   }
\label{figcumulative}       % Give a unique label
\end{figure}

Fig.\,\ref{figcumulative} shows the cumulative distribution versus \Porb\ for the classes of objects we have been examining. The LMXB and pulsar data are taken respectively from the Ritter LMXB catalogue and from the  Australian Telescope National Facility (ATNF) pulsar catalogue\footnote{https://www.atnf.csiro.au/research/pulsar/psrcat/}. No completeness study is considered here, but the catalogue of the high luminosity non transient LMXBs is complete, and it is improbable that the shape of the distribution is altered when the pulsars database becomes more complete. 

%On panel (a) we show \Porb$<15$\,hr, that is the systems whose secular evolution is due to AML. We plot here the number counts for MSP and LMXBs, and the number counts divided by 10 for the CBs. Panel (b) shows the whole cumulative fraction versus the logarithm of \Porb\footnote{the longest \Porb=8578\,d binary MSP J2032+4127 is excluded from the figure.}, and the total numbers in the samples are given in the figure labels.  

In Fig.\,\ref{figcumulative}a (\Porb$<15$\,hr) we see  that neither the LMXB nor the MSP distributions show clear indications of  the minimum period and period gap, well recognized  in the CB distribution. The scarcely populated tail of the distribution of CBs, composed by binaries whose donor is a low mass WD (the AM CVn binaries), is in percentage much more populated in the LMXB sample. On the contrary,  {\it there are no binary MSP below $\sim$2\,hr}, so it is clear  that the LMXB evolution below \Pbif\ has no obvious correspondence with the evolution of CBs in the same \Porb\ range (see Sect.\,\ref{secular}).

In Fig.\,\ref{figcumulative}b we see that there are only $\sim$5\% of  CBs above 8\,hr period 
%(52 ot of 934 CBs in the last version of Ritter \& Kolb catalogue, see also Fig.\,\ref{ritter}), 
and only few objects above \Pbif.
%\footnote{The most famous of such objects is the former nova GK\,Per, at \Porb\ of two days)}. 
On the contrary, at \Porb$>$\Pbif there are several LMXB, and most of the MSP (for this latter category, this is simply due to the fact that these systems are endpoints of the evolution). 
%%%%% EARLY FIGURE 5 %%%% figcumulative
%%%%% FIGURE 3 %%%% figcumulative
 
The total number of MSPs in the catalogues is  2.5 times that of LMXBs. When completeness studies are performed on both samples, the long-standing problem of discrepancy between the birthrates emerges.  First Kulkarni \& Narayan \cite{kulkarni1988} noticed that the birthrate of galactic LMXB is  10--100 smaller than the birthrate of MSP. This latter is now estimated as $B_{MSP} \sim 2.5 \times 10^{-6} yr^{-1}$ \cite{levin2013}. The birthrate of LMXB can be simply computed as $B_{LMXB} \sim N_{LMXB}/\tau_{LMXB}$, as the LMXB sample does not suffer of any significant incompleteness factor (greatly affecting, on the contrary, the  MSP number counts). As $ N_{LMXB} \sim$100, to obtain the same birthrate of MSP, we need a $\tau_{LMXB} \sim 4\times 10^7$\,yr, much shorter than  the typical MB AML ($\sim5\times10^8$yr) or GR AML ($\sim 10^9$yr) timescales.   \\
Solution proposed for this discrepancy are indeed variegate\footnote{We quote here that the presence of single MSPs in the field (and the numerical overabundance of MSPs with respect to their supposed progenitors interacting X-ray binaries) has risen the suspect that some MSPs may have been born directly with their present low magnetic field and rapid spin \cite{miller2001}}, including the role of AIC in the direct formation of MSPs, without passing through the LMXB phase \cite{nomoto1987}, the direct formation of NS having a low magnetic field \cite{miller2001} and the role of `evaporation', the ablation of low mass donors through the wind excited by the high energy pulsar radiation impact, see Sect.\,\ref{ablation}). %(\cite{becker1995}
In Sect.\,\ref{xraycycles} we discuss that the irradiation cycles, plus a late final ablation, can be a reasonable explanation for this puzzle.

%%%%% EARLY FIGURE 6 %%%% fig6
%%%%% FIGURE 4 %%%% fig6
\begin{figure}[t]
\begin{center}
\includegraphics[scale=0.40]{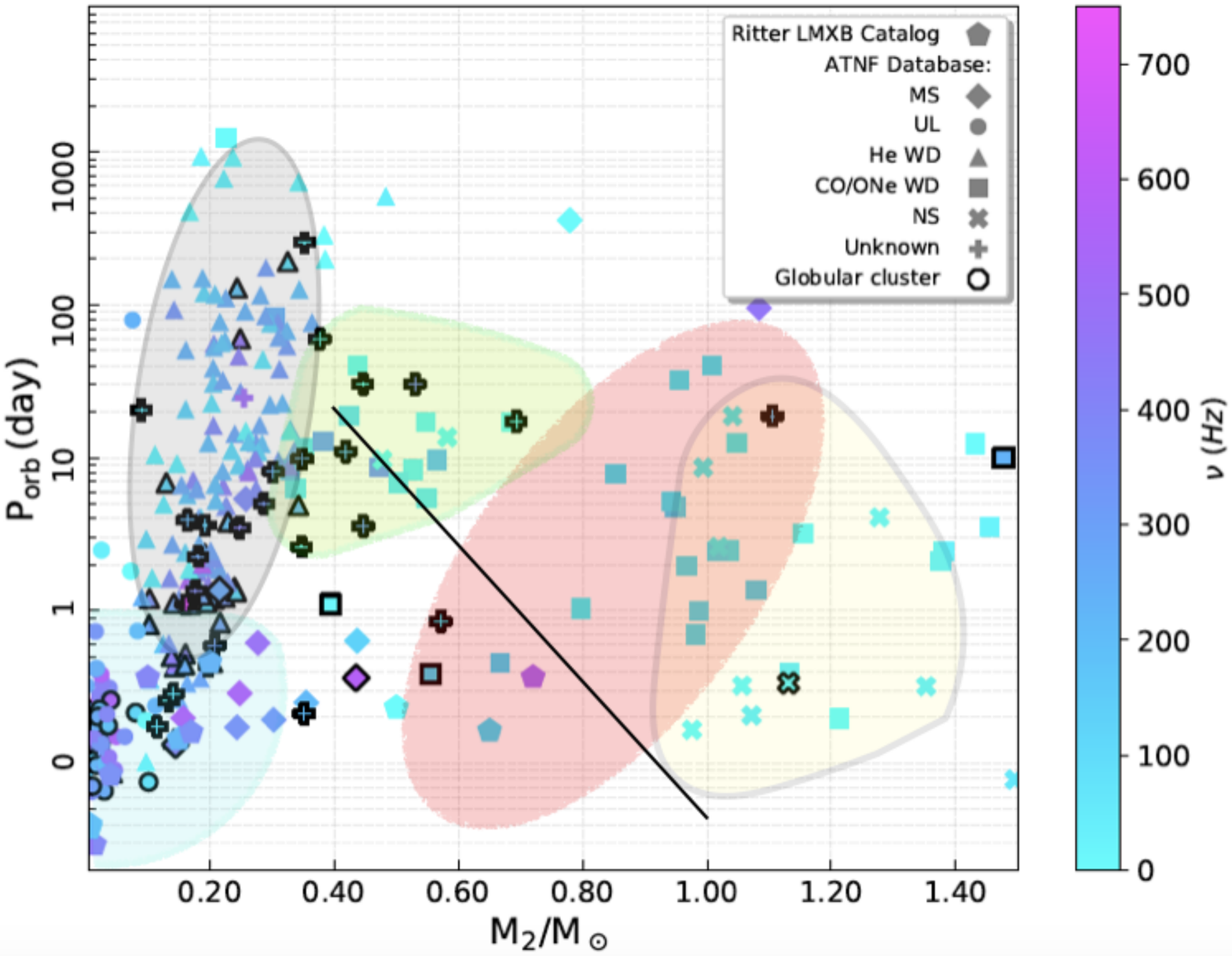}
%\vskip -60pt
\caption{\Porb\ versus average companion mass  of LMXBs \cite{ritterkolb2003} and MSPs from the ATNF Pulsar Catalogue 2018 are shown with different symbols according to their nature. The colors correspond to the pulsar spin, according to the scale on the right. The diagonal line is the relation %\ref{pmc} 
minimum \Porb(\Mc) for case B evolution \cite{taam2000} leading to systems with CO WD companions.
The outcomes (\Porb$\gtrapprox$\,1\,d), in which the companion is a helium WD, are highlighted in grey. The green--yellow shape highlights the location of remnants of Case A and early case B evolution of IMXBs; the red region highlights the remnants of IMXB case C evolution, including a CE phase. NS--NS remnants, higlighted in yellow, result from HMXBs. 
%For the low--period vs. low masses region (cyan) see Fig.\,\ref{Fig.8}.
}
\end{center}
\label{fig6}       % Give a unique label
\end{figure}
 
\section{The  \Porb versus \Mdon\ plane as a tracer of the NS to MSP evolution}
\label{sec:5}
MSP formation follows a variety of binary channels (see, e.g. \cite{batta1991, taurisvdh2006}), and in most cases the binaries are now detached. 
We make a rapid excursion on the MSP systems where the companion is a remnant of the evolution of intermediate and high mass binaries, 
%For these systems, the recycling is definitely over, and they represent the end--point of the binary evolution to the MSP stage. We can gain information on the modalities of mass transfer, mass and angular momentum loss during the semi--detached evolution, and on the critical and uncertain CE phase. \\ 
%We also 
and consider the MSP remnants of  the case B evolution of LMXB.
%, which at a first look appears the same as the standard evolution described in Sect.\,\ref{bifurcation}.
%and the case A evolution (\Porb$\lessapprox$\,1\,d), covering hydrogen rich donor masses $\lessapprox$\,0.5\msun, down to planetary--type masses. 
%\subsection{The \Porb versus \Mdon\ plane as a tracer of the NS to MSP evolution}

 Once we accept the concept that MSPs are recycled, generally old or very old neutron stars, the plot of companion mass (\Mc) versus orbital period (\Porb) provides a snapshot of the evolutionary path through which the neutron star has been accelerated\footnote{We use from now on the term \Mc\ for the NS companion, both for the detached systems and for the still interacting systems. When discussing the secular evolution, in some cases, we will go back to the term `donor' mass \Mdon.}.  
 
\Mc\ is generally not well determined. We know the projected orbital velocity of the NS itself, through the MSP spin doppler effect, and so we have  the mass function of the companion $M_{\rm c}^3 \sin^3 i/(M_{\rm c}+M_{\rm NS})^2$, where $i$\ is the orbital inclination angle. Thus the minimum NS mass \Mc\  can be constrained by assuming a standard value for the NS mass, while an `average' mass is derived  by fixing an `average' inclination ($M_{MSP}=1.35$\msun\ and $i=60^o$\ in the ATN catalog). Anyway the value assumed for the NS mass is itself a lower limit, if we consider that MSP must have accreted 0.1-0.2\msun\ to reach their spin. Strader et al. \cite{strader2019} have determined the average mass of 24 confirmed or candidate redbacks components by optical spectroscopy, and the average NS mass is $\sim$1.8\msun,  consequently their also find larger \Mc's. 

Data with available information are shown in Fig.\,\ref{fig6}, where we also show the  subdivision of different loci in the plane  (inspired by T.Tauris \cite{tauris2011}) for the field MSP.

\subsection{Double NS remnants}
\label{DNS}
The possible evolutionary histories of double NS binaries (DNS)  have been originally studied in view of the importance of the Hulse--Taylor binary PSR B1913+16, and more recently because the merging of the two NS components is a possible source of gravitational waves bursts, as occurred in the LIGO/Virgo events GW\,170817 \cite{abbott2017a} and GW\,190425\cite{abbott2020}\footnote{GW\,190425 has a total mass of 3.4$^{+0.3}_{-0.1}$\msun, 5$\sigma$ larger than the galactic population mean, pointing to a peculiar formation mechanism. Nevertheless, among the possible explanations (see \cite{abbott2020}), the high mass of the primary component points towards recognizing it as a NS which has been subject to non negligible accretion after its formation, that is an MSP.}. 

DNS occupy the region of highest \Mc\ highlighted by the yellow colored region in Fig.\,\ref{fig6}, and are born from high mass binaries evolution. In the figure  we may also notice that the spins of the MSP component are not extreme, meaning that the primary NS is only partially recycled. 

The evolutionary paths to DNS  \cite{batta1991, taurisvdh2006} begins with the
more massive of the two high mass components undergoing a CCSN and leaving a remnant NS, which may receive a high kick at birth. After some time, the secondary, now the more massive component, will evolve off the main sequence out of its Roche lobe, and fully lose the hydrogen envelope in a CE phase. The binary will be then a NS plus a `naked' helium star companion, and may undergo a further phase of mass transfer (the case BB, see, e.g. \cite{dewi2002}),
%Delgado \&Thomas 1981; Dewi et al. 2002; Ivanova et al. 2003; Tauris et al. 2013), 
during which again the donor helium star overfills its Roche lobe. 
In both phases of mass loss after the first CCSN, the binary may appear as a HMXB.
Eventually, a second CCSN will occur, and the DNS will be formed, unless the natal kick of the second SN explosion unbinds the system. 
The sequence of events is quite uncertain, because it depends both on the outcome of (two) CE evolutions, and on the uncertainty in the distribution of natal kicks \cite{tauris2017}. 

A different formation channel  is possible in GCs: the MSP is formed through one of the other binary evolution paths described below, and the remnant donor mass suffers an exchange interaction with another NS in the GC core, resulting in the exchange of the former donor (the lighter star in the triple) with the NS. The three body interaction may be able to eject the newly formed binary to the outskirts of the GC, from which it may go back to the core due to dynamical friction on a timescale longer than 100\,Myr. Such evolution has been suggested to explain the location of the binary pulsar PSR\,B2127+11C, in the GC M\,15, %(and perhaps 1744?24A in Terzan 5) 
far from the cluster core \cite{phinney-sigurdsson1991}. The exchange interaction is also testified by the high eccentricity ($e$=0.681) of this system, and Phinney and Sigurdsson \cite{phinney-sigurdsson1991} argue that similar interactions, in a shorter period system, may lead to ejection of the newly formed DNS from the GC or to NS merging by orbital decay due to GR. 
More recently, Andrews and Mandel \cite{andrews-mandel2019} consider a sub-population of short--period, high eccentricity field DNS %(one is the Hulse--Taylor binary PSR\,B1913+16, having $e$=0.617), 
and notice that they have properties similar  to PSR\,B2127+11C. Rejecting the formation channel of these eccentric DNS in a single binary, they argue that they could have been formed in, and then ejected from, GCs.  
%This scenario provides a pathway for the formation and merger of DNSs in stellar environments without recent star formation, as observed in the host galaxy population of short gamma-ray bursts and the recent detection by LIGO of a merging DNS in an old stellar population.      

% (Phinney and Sigurdsson \cite{phinney-sigurdsson1991}) one of shorter period would have been ejected from the cluster or would have collapsed because of orbital decay by gravitational radiation. Pulsars and pulsar binaries ejected from clusters will contribute to the birth rate of recycled pulsars in the inner Galaxy.

\subsection{Intermediate mass cases A, B or C}
\label{IMtoMSP}
The central region of the data of Fig.\,\ref{fig6}  contains the end--product of the evolution of intermediate mass binaries at different stage of interaction (cases A and B --region highlighted in green-- or case C --highlighted in red
%\footnote{Note that the green region, and the upper part of the red region, when the nature of the companion is not known, may hide LMXBs in a detached stage during the evolution with mass transfer which ends in MSP plus low mass He--WDs. Detachment would be due to illumination cycles \cite{benvenuto2014}.}) 
and characterised by the companion Carbon Oxygen ($M \gtrsim 0.45$\Msun) or Oxygen--Neon ($M \gtrsim 1.05$\Msun). 
%We follow the outline in \cite{tauris2012} and references therein.
We refer the reader to the outline in \cite{tauris2012}  and references therein for an extensive discussion.

%Case A evolution must start from donor masses of 3-5\msun, and leads to final \Porb=5-20\,d. The mass transfer rate is initially huge, well above Eddington' rate, and mostly not accreted on the NS, a short phase of transfer on the nuclear timescale follows, with $\dot{M} \sim 10^{-9} M_\odot/yr$, followed by the phase in which the donor expands to become a giant, at higher $\dot{M} \sim  10^{-8}-10^{-7} M_\odot/yr$, during which about 0.1\msun\ is transferred and the NS can finally be accelerated to MSP.

%In the early case B evolution from donors of 2.5-5\msun\ at starting \Porb$\simeq$3--10\,d, most of the transferred donor mass is lost from the system, while a small few $0.01$\msun\ can be accreted and mildly recycle the NS. 
%The larger is the initial ratio between the donor and the NS mass, the shorter the final period, and the remnant helium star evolves directly into a CO WD. The limiting \Porb\ versus remnant mass is estimated in \cite{taam2000} as
%\begin{equation}
%\log P_{\rm orb~final} (d) \simeq 2.56 - 3.1 M_{\rm c}/M_\odot 
%\label{pmc}
%\end{equation}
%and is shown as a diagonal line in Fig.\,\ref{fig6}.
%If the donor mass is $>$5\msun, a CE phase is unavoidable, probably followed by a merger.

%Case C begins at long \Porb, but the ensuing CE phase does not lead to merger, because the AGB envelopes are weakly bound. The remnant is a CO WD, if the initial masses were below the minimum mass for non degenerate C--ignition (see Fig.\,\ref{fig1}).

%%%%% EARLY FIGURE 7 %%%% fig7
%%%%% FIGURE 5 %%%% fig7
\begin{figure}[b]
%\sidecaption[t]
\vskip -10pt
\begin{center}
\includegraphics[scale=0.33]{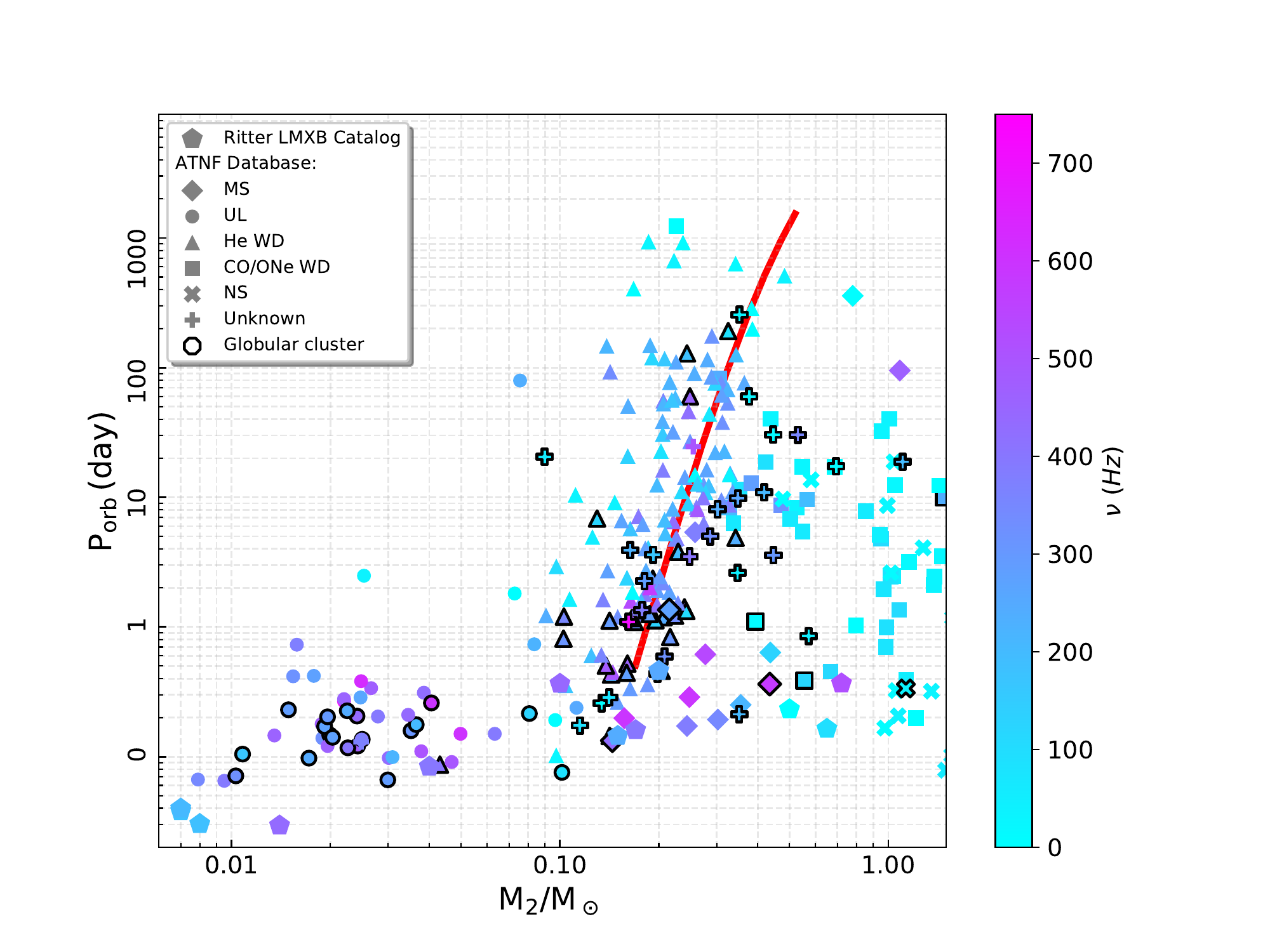}
\hskip -25pt
\includegraphics[scale=0.25]{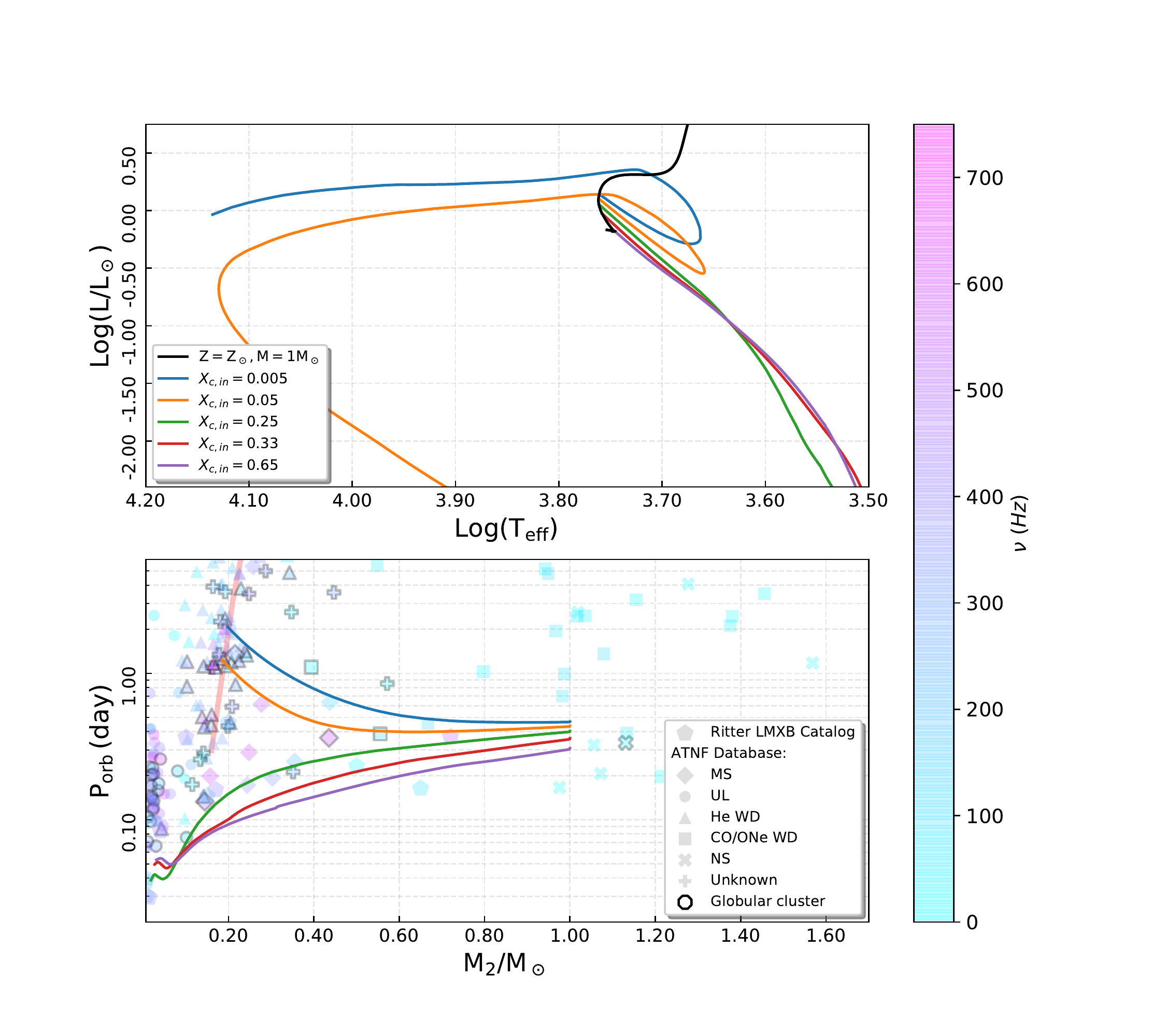}
\end{center}
\vskip -20pt
\caption{In the left panel, the \Porb\ versus log(\Mdon) plot is shown for the lower mass side, where we see well the location of helium WD companions at \Porb$\gtrapprox$\,1\,d. Analytic expression from \cite{lin-rappaport2011} is plotted for masses fro 0.15 to 0.5\Msun.
The right panel shows  typical `classic' evolutionary tracks of the donors in the HR diagram (upper panel) and in the   \Porb\ versus log(\Mdon) plane (bottom).}
\label{fig7}       % Give a unique label
\end{figure}

\subsection{The evolution to MSPs with companion low mass white dwarfs}
\label{mspcaseb}
Fig.\,\ref{fig7} shows the lower side of the \Porb\ vs. \Mdon\ plane. 
%where we see the location of helium WD companions at \Porb$\gtrapprox$\,1\,d. 
The evolution above \Pbif\ at a first look appears to be standard (e.g. \cite{webbink1983}). 
Giants at M$\lesssim$2.2\msun increase their radius around the growing of the helium degenerate core. If the donor mass is larger than the NS mass, initially mass transfer proceeds on the thermal timescale and \Porb\ decreases. When the mass ratio is reversed, mass transfer proceeds on the nuclear timescale and is ideal to accelerate the pulsar, which will emerge when the donor, lost the majority of its hydrogen envelope, contracts to become a helium WD. Thus the period distribution of MSP with He--WD companion, from a few to hundreds of days, represents the frozen endpoints of binary evolution. 
The existence of a He-core mass {\it vs.} radius relation ---almost independent of the total mass--- implies, that there will be a well defined period at which the donor detaches from the Roche lobe for each core mass, and thus there is a relation between the WD remnant mass and \Porb (see \cite{lin-rappaport2011} and the red line in Fig.\,\ref{fig7} for a recent expression). 

%Alessandro suggested
%RIVEDERE huntsman pulsars
A class of systems progenitors of these MSP with He-WD companions  may be the recently discovered systems 1FGL J1417.7--4407 (\Porb=5.4\,d, \cite{swihart2018}) and 
PSR J1306--40 (\Porb=1.096\,d \cite{swihart2019}), dubbed `huntsman' pulsars (referred to the classification of companions of MSP in different classes of spiders (see Sect.\,\ref{secular}). We discuss in Sect.\,\ref{scarcely} the ambiguity in the evolution of such systems, with reference to the similar case of PSR\,J1740-5340.
 
%The region of companions having \Mc$\sim 0.1- 0.3$\msun\ should be populated in the  lower \Porb\ range ($P_{\rm orb} \lesssim$0.3\,d)  by systems beginning mass transfer from not yet evolved donors, while the binary MSP clearly remnants of  case B evolution should be at \Porb$\gtrsim$\ few days. 
From the `bifurcation' of the evolutionary tracks, either towards long \Porb\ and degenerate companion, or to short \Porb\, for core H--burning companions,  we expect a dearth of systems in the region  0.3\,d$\lesssim$\Porb$\lesssim$ few days. This dearth of system is not observed  (see Sect.\,\ref{scarcely}). 
\subsubsection{The `period gap' of MSP with He--WD companions}
Taam, King and Ritter \cite{taam2000} studied the evolutionary channels leading to MSP with remnant companions He or C-O WDs and found out that there would be a natural separation of systems between long \Porb\ ($\geq 60$d) or short \Porb ($\leq 30$d), giving an explanation of the lack of systems at 23$\leq$\Porb(d)$\leq$56. The separation would be due to the fact that, for comparable initial \Porb, low mass case B evolution leads to long final \Porb, while higher mass case B leads to shorter final \Porb.  \\
An alternative solution \cite{dantona2006} relates the gap to the `bump', the stage at which the low mass giant donor would reach the chemical discontinuity left by the maximum deepening of convection and suffer a temporary contraction. The contraction detaches the system and the MSP is activated and may lead the system into radio-ejection, not allowing mass accretion when the giant fills again its Roche lobe. Then the mass is lost from the system carrying away a specific AM which is {\it at least} the AM at the lagrangian point L1,  and the binary final period will be shorter than the final period for conservative evolution. As the bump corresponds to \Porb$\sim$17\,d, the final periods will be 20--24\,d. Binaries initiating mass transfer when the giant has already passed through the bump will populate the range above 60\,d. The prediction of this model is that the  magnetic field of long period MSP should be lower than for the short periods. This model may hold only if the accretion is not yet subject to radio--ejection before the bump \cite{shao-li2012}. 
%%%%%WARNING: RIMESSO QUESTO PARAGRAFO SOLO PER EDIZIONE ASTROPH

%\subsubsection{Standard evolution is not the whole story}
%Focusing on the simple scheme depicted in the lower right portion of Fig.\,\ref{fig7}, 

%As \Porb\ increases during the LMXB evolution, and the final periods are longer than a day, the MSP is not close enough to influence the companion when mass loss ceases, and the systemic angular momentum losses are irrelevant. 

\section{Short period, low-mass companion systems: the mixed bag}
\label{secular}
\begin{figure}[t]
\vskip -50pt
\begin{center}
\includegraphics[scale=0.5]{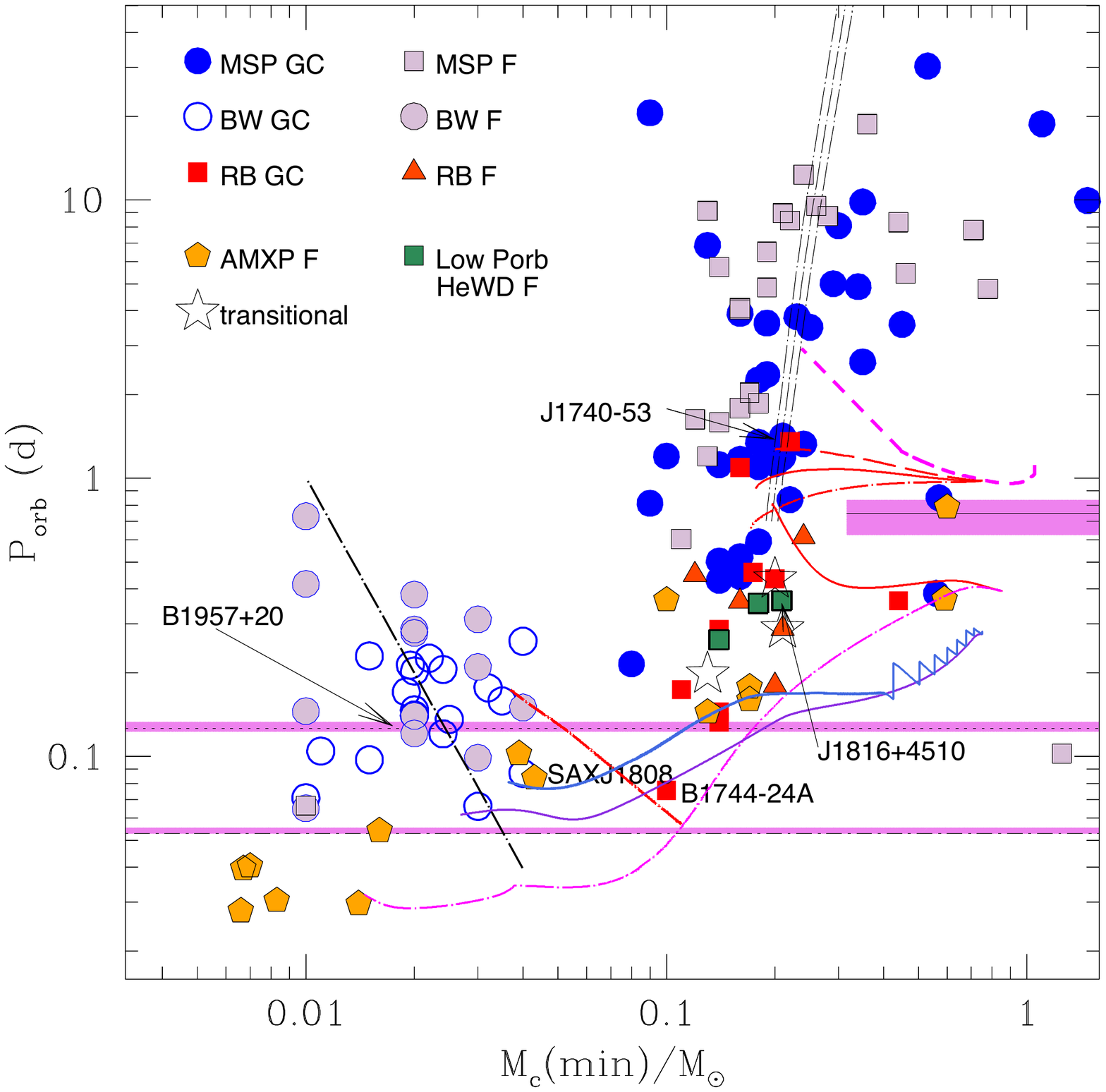}
\end{center}
\vskip -80pt
\caption{Data (Patruno's list), in the minimum \Mc\ versus   \Porb\  plane. \Pbif, the 3\,hr period gap, and the \Pmin=76\,m of CBs are highlighted in pink. Most AMXPs (yellow pentagons) are at \Porb$<$\Pmin. 
Grey squares (GC) and blue dots (field) are standard MSP with (mainly He-) WD companions, and three green squares mark the lowest \Porb\ MSP with He--WD \Mc\ in the field (see \cite{istrate2014}).
Red squares and triangles are the redbacks, and the open stars show  transitional MSP.
Above \Pbif, we show three different evolutions starting at \Mdon=1.05\Msun, in case A, but with an advanced consumption of the core hydrogen. The violet dashed curve and the red lines represent evolution with increasing specific angular momentum losses associated to the mass lost from the system when radio--ejection sets in. Shorter final \Porb\ corresponds to stronger specific AML. Below \Pbif, starting from 0.85\Msun, the dash-dotted magenta line is a standard binary evolution starting in late case\,A, with no illumination and reaching a minimum period well below 76\,m, as the hydrogen content of the core is very small. The violet and blue lines are taken from \cite{tailo2018} and represent a standard case A evolution (no illumination) and an evolution with illumination $\epsilon_X=\epsilon_{MSP}= 0.01$. This latter passes through the location of the AMXP SaX\,J\,1808.4-3658. 
The diagonal dot-dashed lines are an evolution starting from a thermal equilibrium configuration at  $\sim$0.12\Msun\ and including a strong MSP illumination (red line, see text) and a typical evolution with evaporation (black line, from \cite{benvenuto2015bw}).
%The blue evolution represents a case in which the MSP illumination is very large (see text).
}
\label{fig8}       % Give a unique label
\end{figure}

%In Sect.\ref{mspcaseb} we have seen that the region of low \Mc\ and \Porb\ larger than a few days is mostly occupied by He-WD remnants of the evolution to the MSP stage. \\
Overall,  the evolution starting below \Pbif\ results to be the most interesting in the \Porb--\Mc\ plane, because it also hosts systems which are not necessarily at the end point of their orbital evolution, and contains both LMXB (or AMXP) and MSP systems.  
We show in Fig.\,\ref{fig8} the  \Porb\ vs. $M_c$\ data with their identification, from A. Patruno's catalogue\footnote{www.apatruno.wordpress.com/about/ millisecond-pulsar-catalogue/}. 
%We mark the \Pbif $\sim$15-20\,hr band, the 3\,hr period gap high boundary and the \Pmin=76\,m boundary. 
  %\Porb$\sim 15-20$\,hr (close to \Pbif) and the region  \Porb$\sim 7-15~hr$).  
In spite minimum \Mc\ is plotted, there is no doubt that the true $M_c$\ at any given period is {\it smaller } than the donor masses of CBs. 
%%%%%EARLY FIGURE 8 %%%% figpatruno
%%%%% FIGURE 6 %%%% figpatruno
Overall, this region of the \Porb--\Mc\ plane includes:
\begin{itemize} 
\item{\bf AMXP}: accreting millisecond X--ray pulsars. They include MSPs with shortest \Porb. Companions may be H-rich, He- or CO- WDs \cite{patrunowatts2012, campana-disalvo2018}.
\item{\bf Radio MSP binaries}: in these (detached) systems the companion may be a H--burning dwarf or a He-WD. These latter companions are also found at \Porb$<$\Pbif. In many system the companion is not yet identified.
\item{\bf Redbacks}: the red dots (GC) and triangles (field) in Fig.\,\ref{fig8} mark the systems called `redbacks' \cite{roberts2013},  showing extended radio eclipses, associated with circumbinary material,  and $0.1 \lesssim M_c/M_\odot \lesssim 0.7$\ \cite{strader2019}, on almost circular orbits with periods $0.1$\,d$ \lesssim$\Porb$\lesssim$1.37\,d (note that some redbacks are at \Porb$>$\Pbif).
\item{\bf Transitional MSP}: a class of redbacks switching from the AMXP to the radio MSP stage \cite{papitto2015} on timescales of months.
%Note again that the companions have smaller \Mc\  than predicted by standard case A evolution.
%, but shorter periods than the companions white dwarfs expected from case B evolution. 
%The redbacks also include the recently discovered class of `transitional' objects, switching from %Redbacks are found both in the Galactic disk and in GCs (Fig.\,\ref{fig8}).
\item{\bf Black widows} are defined as the family of pulsars in a \Porb\ range similar to the range of  redbacks, but with companions having \Mc$\lesssim$0.05\msun. 
\end{itemize}
%Still keeping in mind that the masses plotted in Fig.\,\ref{fig8} are the minimum masses, there appears a gap in $M_c$\ at   $0.05 \lesssim M_c/M_\odot \lesssim 0.1$.\\
To explain this variety of systems, the  evolution of the donor must include physical inputs due to the NS (and MSP, when accelerated) nature of the accretor, the main reason why the period distribution of LMXB and MSP binaries at \Porb$<$\Pbif\ is very different from the `standard' CB evolution (Sect.\,\ref{comparison-classes} and Fig.\,\ref{figcumulative}). 
When the NS is very close to the companion, the donor suffers from different kinds of irradiation and its structure changes accordingly. Other phenomena as the ``radioejection" (\cite{burderi2002}, see Sect.\,\ref{submsp}) can also be at the basis of the modified evolution.

\subsection{The radius reaction when a source of irradiation is present}
\label{radvarillum}
The total stellar radius derivative in Eq.\,\ref{der1} includes the terms independent on the mass loss, and due to nuclear evolution and thermal relaxation. There is another powerful way of radius change, whose effect is important when the donor has a convective envelope, and it is the `irradiation' or `illumination' term. 
%The modalities by which the presence of an external power source affects the stellar structure have been dealt with in several ways, but the main idea behind these efforts was to understand whether the star would reach a new equilibrium radius such as described in a seminal paper by Podsiadlowski  \cite{podsiadlowski1991}.\\
We write explicitly:
\begin{equation}
\left( \frac{\partial \ln R_d}{\partial t} \right)_{\dot{M}=0} = \left( \frac{\partial \ln R_d}{\partial t} \right)_{\rm{th. rel.}} +  \left( \frac{\partial \ln R_d}{\partial t} \right)_{\rm{ill}}
\label{drdtill}
\end{equation}
(the negligible term $ \left( {\partial \ln R_d}/{\partial t} \right)_{\rm {nuc}}$ is dropped).
The importance of the last term was first shown in \cite{podsiadlowski1991},  addressing the change in the structure of stars with a convective envelope immerged in the X-ray radiation field of the accreting NS. Afterwards, the problem received attention by considering the evolution of the companion of MSP irradiated by the MSP power \cite{dantona1993} or accreting NS irradiated by the X--ray flux \cite{dantona1994, hameury1995, benvenuto2014}, with the main attention focused on the onset of mass transfer. 

When the donor, having an unperturbed luminosity $L_*$, finds itself immersed in a "heating" bath of luminosity $L_h$, the stellar surface suddenly can not  emit $L_*$, so its
structure (its \teff) must adjust so that the star can emit {\it both} the intrinsic and the heating luminosity:
\begin{equation}
L_{tot}= L_*+ L_h = 4 \pi R^2 \sigma T_{\rm eff}^4
\end{equation}
where
%\begin{equation}
$L_{h}= 4 \pi R^2 \sigma T_{\rm b}^4$\
%\end{equation}
defines the temperature of the radiation bath $T_{\rm b}$. 
This is not a problem for the stars having radiative envelopes, as they can adjust easily the temperature gradient in the external layers to do it, and, in fact, illumination is totally irrelevant for donors of M$\gtrsim$1.5\msun. But, for smaller masses, the stellar envelope is convective, and throughout the envelope the temperature gradient is adiabatic, and dictated by thermodynamics (apart from the upper over-adiabatic surface layers). Thus the surface perturbation is immediately felt {\it at the bottom of the convective envelope}. Starting from a configuration of thermal equilibrium ($L_*=L_{nuc}$), the star reacts approximately only on the thermal timescale of the convective envelope itself, 
%\begin{equation}
%\tau_{\rm conv}=\left( \frac{\partial t}{\partial \ln R_d} \right) \simeq \frac{3}{7} \frac{GM_dM_{\rm conv}}{R_dL_{\rm nuc}}
%\label{tauconv}
%\end{equation}
%The timescale is
longer for smaller masses having deeper convective envelopes ($\sim 10^8$yr for a 0.5\msun), but in any case long enough that a ``fully bloated" configuration (those studied in \cite{podsiadlowski1991}) can not be reached. 

A better physical approach to describe the irradiation comes by considering that an irradiation flux $F_{\rm irr}=L_{\rm h}/(4\pi R^2)$\ is deposited  below the photosphere. The illumination, which in principle affects at most half of the donor surface, is thus distributed on the back side of the star on a timescale much shorter than any of the evolutionary timescale into play, and mainly shorter than  \tauconv. The ways in which the flux can be distributed and becomes symmetric has been considered by several studies, notably in \cite{hameury1993, vilhuergma1994}. The nature of the illumination source is relevant, because it determines the depth at which the energy flux penetrates into the envelope: the depth is larger for a harder X-ray flux, but even a soft X-ray spectrum allows $\sim$10\% of flux deposited in the adiabatic part of the convective envelope \cite{vilhuergma1994}, and can be circulated to the dark side on a timescale close to the sound speed. When the deposition density becomes a few tens of g cm$^{-2}$,  the effect on the radius derivative becomes close to that of symmetric illumination \cite{hameury1995}. \\
When the problem was initially approached \cite{vilhuergma1994}, the idea about MSP illumination was that a role was played by the relativistic e$^+$-e$^-$ pairs and $\gamma$\ ray photons present in the beam composition. The FERMI LAT observations have shown that the $\gamma$-ray emission power from MSP may be a considerable fraction (10-90\%) of the spin down energy \cite{abdo2013}. Thus the MSP illumination has the potential to be much more symmetric and effective than the illumination from X--rays during the accretion stages. \\
The roles of illumination by the X--rays in a LMXB, or by the pulsar radiation in a MSP are indeed very different as we are going to discuss.
%\begin{equation}
%\left( \frac{\partial \ln R_d}{\partial t} \right)_{ill} = \left( \frac{\partial \ln R_d}{\partial t} \right)_{X} + \left( \frac{\partial \ln R_d}{\partial t} \right)_{MSP}   
%\end{equation}    

\subsection{The consequence of X-ray irradiation: mass transfer cycles}
\label{xraycycles}
Independent from the `simmetry' of illumination, the activation of the NS as an X--ray source is such that it affects at least a fraction of the donor surface, alterating the ways in which the stellar luminosity (equal to 4$\pi R^2$ times the flux, in standard conditions) is emitted, if about a half of the 4$\pi$\ solid angle is blocked by the X--rays. This will affect the radius and the mass transfer at least in a transitory ways. \\
%
%The presence of a perturbation depending on the mass transfer rate (in the case of X--ray illumination) modifies the mass transfer and leads to evolutionary cycles of mass transfer followed by detachment phases.
At first the donor is in thermal equilibrium. As soon as the Roche lobe approaches to the stellar radius, mass transfer begins, following Eq.\,\ref{massloss}, and most of the gravitational energy liberated by mass accretion on the NS will be emitted in the X--ray region of the spectrum, $L_X$.
%\begin{equation}
%L_X=L_{acc}= \frac {GM_{NS}\dot{M}}{R_{NS}} \simeq 1.2\times10^{36} \left[\frac{M_{NS}}{1.4M_\odot}\right]
%\left[\frac{10 Km}{R_{NS}}\right] \left[ \frac{\dot{M}}{10^{-10}M_\odot/yr}\right]
%\label{LXacc}
%\end{equation}
A fraction $\epsilon$\ of the intercepted power of the X-ray luminosity at the surface of the donor having radius $R_d$, in the binary having separation $a$\ will be the illumination power:
%\begin{equation}
%  L_{intercept}= \left(\frac{R_{d}}{2a}\right)^2 \times L_X
%\end{equation}
%Generally, only a small fraction of the intercepted X--rays (corresponding to the highest energy tail of the spectrum) will be able to affect the donor evolution. So 
%The uncertainty in how effective will be the illumination, both in terms of symmetric  `radiation bath', or as a partial block to the stellar flux radiation, can be parametrized 
%By taking a further fraction of the intercepted power, we can parametrize the uncertainty in the efficiency of illumination:
%\begin{equation}
%  L_{ill}= \epsilon \times L_{intercept}
%  \end{equation}
\begin{equation}
 L_{ill}= \epsilon \times \left(\frac{R_{d}}{2a}\right)^2 \times L_X
 \end{equation}
For typical conditions of LMXB systems, very small $\epsilon$\ is enough to provoke cycles of mass transfer followed by epochs of detachment. The cycles depend on the fact that the average \Mdot\ through the secular evolution is eventually due to the systemic AML, so if \Mdot\  becomes larger than the average, at some stage, because of the additional radius increase due to illumination, it must be reduced in the following stage. The illumination cycles in CBs are due to the (much weaker) UV irradiation of the accreting WD: these have been formally studied by King et al.  \cite{king1995cycles}, showing that they induce a modulation of the mass transfer rate around the rate dictated by the systemic AML. The effect can explain the differences in mass transfer found for CBs having similar orbital periods. 
%%%%% EARLY FIGURE 9 %%%% onset06
%%%%% FIGURE 7 %%%% onset06
\begin{figure}[t]
%\sidecaption[t]
\begin{center}
\includegraphics[scale=0.18]{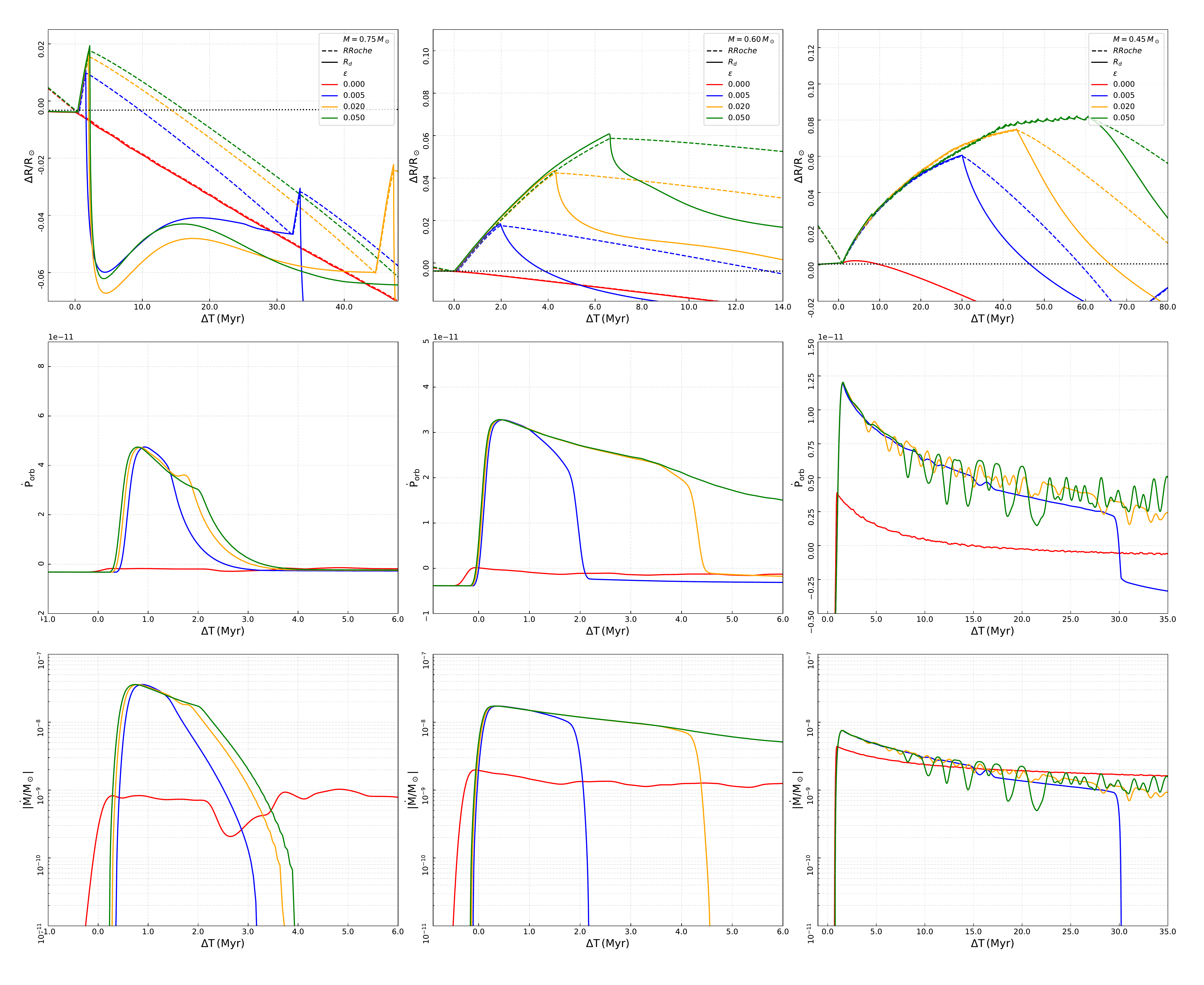}
\end{center}
\caption{Top panels: radius evolution during the first cycles of mass transfer in binaries containing a NS and a main sequence donor of 0.75\msun\ (left panel), 0.6\msun (center) and 0.45\msun (right). Note the different timescales in the abscissa. Mass transfer is computed following the non stationary phases by Eq.\,\ref{massloss} in Sect.\,\ref{mdotdot}. and the mass loss rate is shown in the bottom panels, in an enlarged scale which covers the first cycle of mass transfer only.
This same time scale is used in the Central panels, plotting the derivative of the orbital period. 
The red line shows the evolution driven by AML only, not considering illumination. The radius decreases and the \Pdot\ is negative during the whole evolution. The other lines show the evolution for different fractions of illumination $\epsilon$ from 0.5 to 5\%. The maximum radius expansion increases with $\epsilon$, and the system detaches almost suddenly when the radius increase due to illumination is overcome by thermal relaxation. The dashed lines represent the Roche lobe evolution, which closely follows the radius expansion until detachment, while the following contraction occurs on the timescale of AML, until the stellar and Roche radius meet again and the new cycle begins. Of course, the interval of detachment is longer when $\epsilon$\ is larger. Note that the {\it maximum} \Pdot\ is independent on $\epsilon$.
}
\label{onset06}       % Give a unique label
\end{figure}

\subsubsection{Mass transfer cycles and the redbacks}
%\label{xraycycles}
The cycles of mass--transfer due to X--rays illumination in LMXBs are particularly important, as they may be relevant to understand: 
After early computations \cite{hameury1993, dantona1994, hameury1995}, recently the evolution of LMXBs with illumination has been afforded in \cite{benvenuto2014, benvenuto2015bw, benvenuto2017,tailo2018}. 
% Alessandro
%In \cite{benvenuto2017, tailo2018} it is shown that the cycles provide a good explanation of the redback MSP stage, interpreted as each of intermediate phases in which there is no mass transfer, between two mass transfer (LMXB or AMXP) stages.

Eq.\,\ref{ddotmloss} is modified by illumination, as there is a further perturbative factor for the stellar radius:
\begin{equation}
\ddot{M}= \dot{M} \frac{R_d}{H_p} \left[  \left( \zeta_S - \zeta_{RLd} \right) \frac{d \ln M_d}{dt} + \left( \frac{\partial \ln R_d}{\partial t} \right)_{ill} +
\left( \frac{\partial \ln R_d}{\partial t} \right)_{th.rel.} + \left( \frac{\partial \ln R_{RLd}}{\partial t} \right)_{\dot{M}=0} \right] 
\label{ddotmlossill}
\end{equation}
The term $\left( {\partial \ln R_d}/{\partial t} \right)_{ill}$\ is positive  and acts to enhance the mass transfer rate, contrary to the relaxation term $\left( {\partial \ln R_d}/{\partial t} \right)_{th.rel.}$, negative for this kind of donor. The Roche lobe will follow the radius increase, and the orbital period will increase. 
%When the thermal relaxation prevails, the radius will contract within the Roche lobe and mass transfer will cease. 
In the first stage, \Mdot\ increases, and so increases $L_{ill}$\ while the radius reacts on the thermal timescale of its convective envelope. The larger is the donor mass, the smaller is the convective envelope extension ---and the thermal timescale at its bottom--- and the faster is the radius increase and the larger the \Mdot\ peak.
The limitation in the mass transfer rate is not due to  the impinging radiation level ($\epsilon$), but only to the interplay between the radius expansion due to illumination and the thermal relaxation. As soon as the radius begins to decrease due to the thermal relaxation of the envelope, \Mdot\ decreases, together with the illumination source (proportional to \Mdot), and the radius tends to go back to its thermal equilibrium radius. In the X--ray phase the period has increased due to the anomalous mass transfer, so that the donor finds itself within the Roche lobe and mass transfer stops until the AML brings it back into contact. 

The onset of mass transfer for different initial masses and different fractions of irradiation due to the X--ray luminosity are shown in the Figures \ref{onset06}. The larger is the donor mass, the longer is the interval between the LMXB phases and the higher the peak of mass transfer reached. The dominant factor in the play is the thermal timescale of the convective envelope, which increases with decreasing the mass.\\
The orbital period derivative  \Pdot\ measured in LMXBs is not directly linked to the secular evolution (that is to the angular momentum losses which are the primary drivers of standard evolution) but {\it to the phase at which we are looking at the system, along the cycles of mass transfer}. 

The cyclic evolution is unavoidable for all masses which have convective envelopes, thus $M \lesssim 1$\msun.  In the larger masses the cycles lead to super--Eddington mass transfer rates, and only a minor fraction of the mass lost by the donor can be accreted. The high mass loss means that a few cycles of LMXB will be sufficient to reduce the mass quickly, and the probability to catch an LMXB system with a donor mass in the range $0.4 \lesssim M/M_\odot \lesssim 1$\ is small.  Thus a reason why the donors have smaller mass than the values found for donors in CBs at similar \Porb\ may be linked to the occurrence of super--Eddington cycles of mass transfer.\\
In the range $M/M_\odot \lesssim 0.4$\ the cycles become longer. Nevertheless, the mass transfer remains unstable and thus the system is active as LMXB only for a fraction of the AML timescale. From Fig.\,\ref{onset06}, considering the shorter recurrence between cycles ($\epsilon$=0.005), the LMXB phase is 
$ \sim $1/15,  1/10 and 1/4 of the entire cycle, thus fully solving the birthrate problem.
\\
In the case displayed, and in \cite{tailo2018}, each cycle spans from 30-50\,Myr (0.75\msun) to 20-90\,Myr (0.6\msun) to 70-140\,Myr. This kind of cycles is compatible with the positive \Pdot's found for some AMXPs.\\
In particular, the   \Pdot=3.6$\div 3.8 \times 10^{-12}$\ \cite{sanna2017} measured during the outbursts in SAX\,J1808, can be reproduced by standard evolution including X--ray irradiation and MSP irradiation with $\epsilon=0.01$ \cite{tailo2018}. %SEE ALSO  XX.
\\
The cyclic evolution also accounts for the higher frequency of systems at $0.1 \lesssim M/M_\odot \lesssim 0.2$\ . In fact, only $\sim$10\%  of the total secular evolution time is spent at masses $\gtrsim$0.2\msun, %(Fig. \ref{f2tailo18}), 
the evolution slows down at smaller \Mc\ because the detachment periods become longer. 

Summarizing, the X--ray cycles may be relevant to understand:
\begin{description}
\item{1-} the difference in the birthrate of MSP and LMXBs: the birthrate of LMXBs is reduced, as the timescale of the mass--exchange phases is much shorter than the timescale imposed by systemic AML; 
\item{2-} the faster evolution observed for some LMXBs by measuring their $\dot{P}_{\rm orb}$, in particular the high positive $\dot{P}_{\rm orb}$'s;
\item{3-} the large differences in   $\dot{P}_{\rm orb}$, for very similar systems. 
\item{4-} the cyclic detachment allows to initiate the interaction of the MSP with the companion (MSP illumination, evaporation or radio-ejection) at any stage along the secular evolution, as soon as the NS has been accelerated. There is no need to invoke detachment at, say, the period gap. 
\end{description}

%%%%% FIGURE 10 %%%% f2tailo18
%\begin{figure}[t]
%
%\sidecaption[t]
%\includegraphics[scale=0.2]{./figure/fig2tailo.pdf}
%\caption{The evolution of the companion mass (\Mc) with time for models with different $\epsilon$ of illumination efficiency. The binaries spend only $\sim 20$\% of their lifetime at $M>0.2$\msun. }
%\label{f2tailo18}       % Give a unique label
%\end{figure}

\subsection{The MSP illumination}
%%%%%%%% MODIFICARE
There is a further consequence of the cyclic detachment: when the NS has been accelerated to MSP, the pulsar power may catch on and become the most important source of perturbation of the companion evolution:  the pulsar emission in the $\gamma$\ spectrum \cite{abdo2013} represents a long lived illumination source, and is more `symmetric' and effective than the X--rays during the accretion stage, thanks to its deep penetratation into the donor envelope.
Thus the  X--ray cycles are the way the low mass system is able to recycle the NS companion to millisecond spin periods. \\
The illumination due to the MSP may be the key to understand  the lack of standard MSP binaries below \Porb=2\,hr (other models are available as well). \\
First, the NS has to be accelerated to the MSP phase, so the presence of an illumination due to the MSP depends on the precise effect of the NS spin-up and spin-down mechanisms. The spin-up problem has been afforded numerically in early works in e.g. \cite{muslimov1993, lavagetto2004} and more recently in \cite{tailo2018}. 

According to the Larmor's formula, the pulsar luminosity can be written as:
\begin{equation}
L_{MSP}=\frac{2}{3c^3} \mu^2 \left(\frac{2\pi}{P_s}\right)^4  ~~erg/s
\end{equation}
where $\mu$\ is the magnetic momentum of the pulsar in G$\times cm^3$, $c$\ is the speed of light and $P_s$\ the spin period in seconds. With the typical values of magnetic momentum, derived from the MSP spin down, the MSP luminosity is much smaller than the X--ray luminosity in the typical LMXB regime, resulting to be $10^{32} - 10^{35}$erg/s, to be compared with $L_X \sim 10^{38}$erg/s in systems accreting at Eddington rate.\\
Nevertheless, even a small fraction of MSP illumination is sufficient to allow the donor to remain much more expanded than obtained by the thermal disequilibrium due only to systemic mass loss mechanisms \cite{tailo2018}, and  the minimum period reached by these systems will be much larger than the 80\,min of CBs, depending mostly on the fraction of MSP illumination allowed. 
The results of the computation in \cite{tailo2018}  show explicitly that the important factor to produce long term effects is the action of the MSP spin-down luminosity, even at the low level of 1\% efficiency as in the case shown in Fig.\ref{fig8} (blue  line). X--ray illumination only is  totally ineffective in the long term and the evolution is very similar to the standard evolution (violet in Fig.\,\ref{fig8}).\\
Thus  the role of the X--ray illumination for the secular evolution is almost negligible, but it allows the existence of cycles of mass transfer, so that the system is detached for most of the time, and the much more efficient and symmetric illumination due to the MSP spin down luminosity is the true responsible for the peculiarity of the binary evolution in the MSP binaries.

\subsection{The `evaporation' model: a role for  the black widows stage? }
\label{ablation}
We have seen in the preceding section that at least some systems in the BW region (\Mc$\lesssim$0.05\Msun\ and \Porb$\sim$2--20\,hr) can result from standard evolution from a main sequence donor, with a small degree of illumination in both X--ray and MSP stages. But the BWs are most probably the result of a variety of possible evolutionary paths, and generally `evaporation' is indicated as the dominant active mechanism.\\
The first evidence of the effect of the role played by a close MSP on the binary evolution was the discovery of wind mass loss from the companions of two MSP:
\begin{description} 
\item 1) PSR\,B1957+20  \cite{fruchter1988psr1957+20} at \Porb=9.2\,hr, with an \Mc$\sim$0.025\Msun. The companion eclipses the pulsar for $\sim$50\,m, and delays the pulses for a few minutes before and twenty minutes after the eclipse, indicating the presence of surrounding plasma. 
\item
2) PSR\,B1744-24A (J1748-2446A) in the GC Terzan 5 \cite{lyne1990psr1744, nice1990}, at \Porb=1.8\,hr with variable eclipse duration from1/3 to 1/2 of the orbit. Thus the eclipsing region is much larger than the Roche lobe of the companion having \Mc$\sim$0.1\Msun. This MSP system becomes then a prototype of redbacks.
\end{description}
Does the impinging radiation from the MSP cause a wind mass loss with an associated AML due to the MSP pressure acting on the wind? When these systems were observed,  already a model predicting `evaporation' from a close companion of an MSP had been developed by Ruderman et al. \cite{ruderman1989pulsar}.
%and also applied to LMXB \cite{ruderman1989X}.
 The physics of evaporation presents several big question marks. The efficiency in producing the wind (`wind driving') is the first quantity to be understood (or parametrized). The following problem is how AM is lost from the system during the wind phase, as the period evolution largely depends on this input \cite{tavani1991,ergmafedorova1991,ergmafedorova1992}.
SPH simulations in 2D %(e.g. \cite{tavanibrook1991}) 
and 3D %\cite{tavanibrook1993, tavanibrook1995},
\cite{tavanibrook1993}, led to the proposal of a model of self--excited evolution of LMXB \cite{becker1995}, in which the mass and AML due to the evaporation stimulated further mass loss till the reduction of the companion of the MSP to the planetary mass stage.\\
%Early attempts to parametrize wind mass loss were done to account for the lack of LMXBs below the period gap \cite{vdhvp1988}. 
A `fast evaporation' model was developed by Stevens et al. \cite{stevens1992} to account for the presence of  `planets', companions of a few Earth masses, around pulsars \cite{bailes1991},% wolszczan1991}, 
that  would be formed in the thick disk around the NS, due to the evaporation of the MSP companion. In this study, the authors propose a simple expression for the evaporation:
\begin{equation}
\dot{M}_{\rm evap} = \zeta L_{\rm MSP}  \left(\frac{R_d}{a}\right)^2 
\end{equation}
where $R_{d}$\ and $a$\ are the donor radius and the separation. The normalizing factor $\zeta$\ is estimated by requiring that the pulsar luminosity $ L_{\rm MSP}$\ incercepted (times an efficiency factor $f$) is equal to the kinetic luminosity of a thermal wind with speed equal to the escape velocity $v_{esc}$\ from the stellar surface:
\begin{equation}
 f L_{\rm MSP}  \left(\frac{R_d}{2a}\right)^2 = \frac{1}{2} \dot{M}_{evap} v^2_{esc}
\label{evap} 
\end{equation}
This assumptions is very important for the model, because the brown dwarf donor will inflate ($M \propto R^{-3} $) losing mass adiabatically, $v_{esc}$ becomes smaller and evaporation faster.\\
The evaporation model has been applied in several recent works \cite{benvenuto2012, chentaurishan2013, benvenuto2015bw}. 

%DESCRIPTION OF PROs and CONTRAs follows: mainly the Chen model begins at the period gap and does not give an explanation for the fact that the BW fills almost the Roche lobe!! Benvenuto has something about the INITIAL masses that must be in the limited radng M>1.5Msun. How do we deal with the GC source (Stella)?

Chen et al. (2013) \cite{chentaurishan2013} focus on the bimodal distribution of MSP of low \Mc, between the region of \Mc $\gtrapprox$0.5\msun,  populated by redbacks and the region of \Mc$\lesssim$0.05\msun, populated by BWs. They suggest that in both regions evolution is dominated by evaporation, but at different efficiency. 
%is adopted, by chosing different $f$\ efficiencies. 
The model proposes that the NS has been accelerated by a standard LMXB evolution until the system detaches due to the radius contraction at the period gap. At this point, the donor may recover its equilibrium radius, if $f$\ is small, or begin the evaporation phase, if $f$\ is large. In the latter case, the system will find itself in the redbacks region, in the first case it will populate the BW region. The reason for different values of efficiency may reside simply in the distribution of angles between the orbital angular momentum and the pulsar magnetic axis. This model anyway has to include further hypotheses to account for the redbacks having \Mc\ larger than the typical 0.2\msun at which the period gap would appear.

The \cite{chentaurishan2013}  model is interesting because it propose a unique mechanism both for the redback and BW stage. We will see in Sect.\,\ref{scarcely} a different proposal linked to illumination and radio-ejection for redbacks. Further, we have  seen in Sect.\ref{xraycycles} that the structure of the companion modified by illumination delays reaching of a fully convective structure to below 0.1\msun, and anyway X--ray cycles predict detachment at any stage, independent of the presence or not of a period gap.
Thus it is more probable that evaporation has a role mainly after the binary has already reached a short period during the previous evolution. 

Benvenuto et al. (2012, 2014, 2015) \cite{benvenuto2012, benvenuto2014, benvenuto2015bw}  model the evolution of  black widows through evaporation. The simple prescription in Eq.\,\ref{evap} \cite{stevens1992}  
is adopted, fixing the efficiency at $f=0.1$. The evolution is followed through cycles of mass transfer (see Sect.\,\ref{xraycycles}) until the mass-radius relation of the donor reverses,  at the minimum period (see Sect.\,\ref{sec:4}) and then the evaporation acts to increase the orbital period. Different degrees of hydrogen consumption in the donor are followed. The progenitor of BW, in this scenario, evolves when the initial period of the binary is close to the bifurcation. Benvenuto et al \cite{benvenuto2015bw} show that the range in their models is  \Porb$\sim 0.6 - 1$\,d for \Min=1.5--3\Msun. The starting relatively large initial masses are necessary to reach the observed average density of the companion and the mass ratio of some BW binaries. The partial hydrogen consumption in the core is also consistent with the very hydrogen poor spectrum of the companion of PSR\,J131-3430 \cite{romani2012}. Other BWs have standard hydrogen rich companion spectra, so the BWs appear to be a mixed bag themselves.

We must not forget that the BW binaries might not be the end point of the secular evolution in a single binary.
King, Davies \& Beer \cite{king2003} noticed that BWs are more abundant in GCs than in the Galactic
field, and in particular, GC MSP binary lack the rich population of remnants of case B evolutions of intermediate and high mass present in the field (see Fig.\,
%\ref{fig6} ___ does not work the reference to fig. 4!
4, 
where the systems above the line of minimum \Porb\ for these evolutions are all in the field). Thus they suggested a two--stage formation: first a binary evolution leading to the acceleration of the NS, generally in a case B evolution, leaving a loosely bound WD companion; a second event would be an exchange interaction by which the low mass WD is substituted by a main sequence star (of mass most probably close to the turnoff mass of the epoch of the exchange) leaving a high eccentric orbit. Tides at periastron reduce the eccentricity and, added to further encounters, lead to a closer orbit. The further evolution would be dominated by radio--ejection and lead the system to a BW stage. Field BWs then require the ejection of the system from the parent GC. 

The BW companion can be the same star which spun up the NS only if the detachment occurs at short \Porb, and that GR can bring the MSP and the companion close enough to start evaporation on a timescale shorter than the age of the Universe.\\
%\cite{king2005}. 
An alternate path to the BW region occurs if we consider the effect of MSP illumination  \cite{dantona1993} during the time of action of GR in a detached stage such that the low mass companion has reached thermal equilibrium again\footnote{This case is different from those discussed  in  \cite{tailo2018} ---e.g., the evolution passing through SAX\,J1808 in Fig.\,\ref{fig8}---
where the donor was never able to go back to full thermal equilibrium, thanks to the small X--ray and  MSP irradiation,  at the detachments following the X--ray cycles. In the case here examined, the evolution starts with a companion in thermal equilibrium, and for this reason the effect of the MSP irradiation is larger, see, e.g. the models in \cite{dantona1993} }. If the $L_{MSP}$\ is indeed 10--90\% of the spin down energy, as implied by the gamma ray observations, the donor expands to its Roche lobe and goes on expanding while mass is lost by the system, and \Porb\ increases. In Figure\,\ref{fig8} we show the possible path starting at 0.12\Msun\ and ending at 0.04\msun\ and \Porb$\sim$5\,hr, in the middle of the BW region. The timescale of this evolution will be $\sim 0.5 - 1 \times 10^9$\,yr, if the mass loss is dominated by the companion expansion due to the illumination. If evaporation also is active, the timescale becomes shorter. The advantage of including illumination is that the companion will be always close to Roche lobe filling, and the mechanism will be more effective (see Eq.\,\ref{evap}).
%we show the computation of evolution starting from a donor of 0.12\Msun\ in thermal equilibrium, where the strong pulsar illumination actes on the stellar radius expansion on the very long thermal timescale of the star, so that it fills the Roche lobe and begins its final evolution. Mass loss from the system increases \Porb\ up to $\sim$5\,hr, in the middle of the BW locations. This kind of evolution is entirely due to the   

\subsection{The evolution close to \Pbif}
\label{scarcely}
In Fig.\,\ref{fig8}, the two red squares (redbacks in GCs) just above \Pbif,  are PSR J1748--2446ad, the fastest MSP at 716\,Hz found in the GC Terzan\,5 at \Porb$\sim$26.26\,hr \cite{hessels2006}, and PSR J1740--5340, the partially obscured MSP in NGC\,6397 at \Porb$\sim$32.5\,hr \cite{damico2001}. The radio eclipses in both these systems are of variable length, leading to model the companion as a star losing circumbinary material, as it would occur in the `radio-ejection' model.
%%%%% EARLY FIGURE 11 %%%% doppia per 6397 e redbacks
%%%%% FIGURE 8 %%%% doppia per 6397 e redbacks
%% Near the bifurcation period the nuclear timescale of the secondary and the angular momentum loss time?scale are very close to each other. Therefore, during the binary evolution orbital period changes are insignificant in comparison with the initial orbital period (Tutukov et al. 1985; Ergma et al. 1998). 
\begin{figure}[t]
%
%\sidecaption[t]
\vskip -30pt
\begin{center}
\includegraphics[scale=0.27]{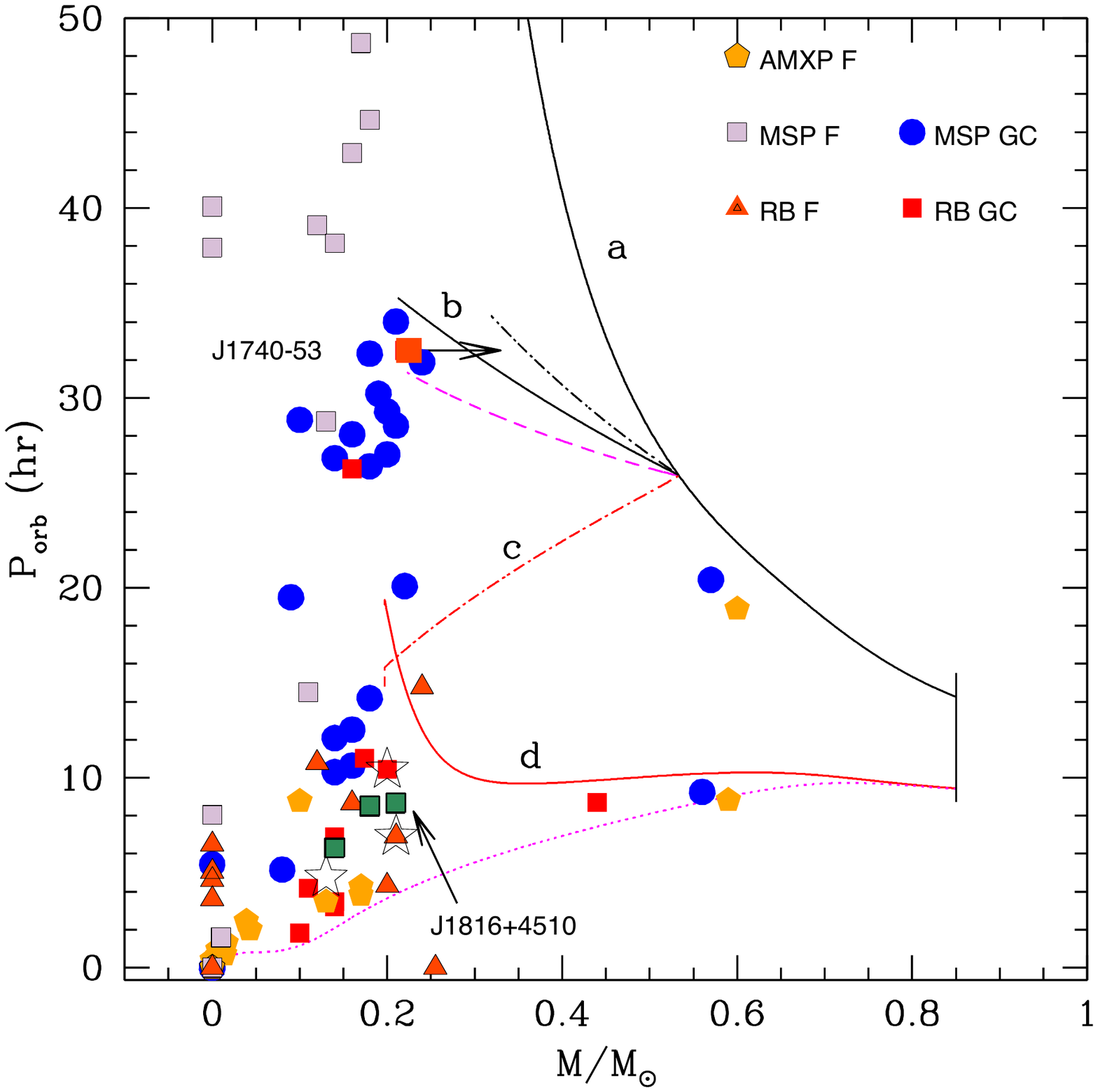}
\includegraphics[scale=0.27]{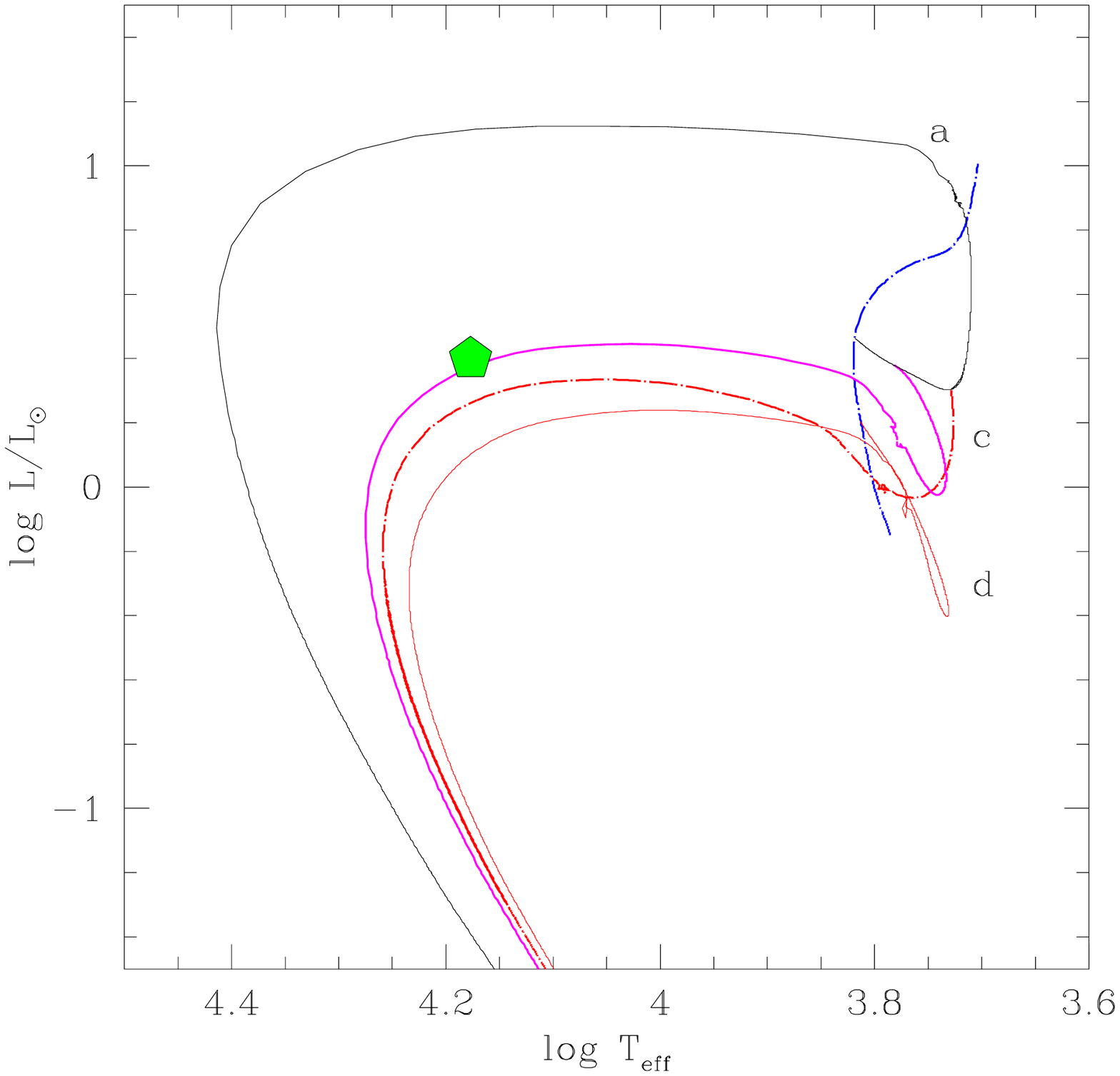}
\vskip -40pt
\caption{Evolution of the companion of PSR\,J1740-5340, whose plausible mass range is limited by the red square (minimum mass) and the black arrow (\Mc$\lesssim 0.35$\msun). The location at the turnoff of the Globular Cluster NGC\,6397 implies an initial mass  $\sim$0.85\msun. Left panel: Standard evolution with radio-ejection begins at P=29\,hr, assuming that the mass is lost with the specific AM at the Lagrangian point (curve a \cite{burderi2002}) and  predicts \Mc$\sim$0.5\msun\ at  \Porb=32.5\,d.  A larger loss of specific angular momentum is needed to fit the system data (e.g. curve b). In the extreme assumption (curve c) that the specific AML is 20\% larger than at the donor location, \Porb  {\it decreases} .
We show also two evolutions starting with a much less evolved stage of the donor of 0.85\msun, with different primary losses of angular momentum.  %The line increasing \Porb\ (d) assumes a low efficiency of magnetic braking.
In the right panel some of the evolutions are shown in the theoretical HR diagram, where also the 0.85\msun\ track is shown (dash dotted blue line). Case c, d of the left panel evolve to the WD stage at luminosities below the main sequence starting location, and may be able to represent the locus of the peculiar companion of PSR\,J1816+4510 (green pentagon \cite{kaplan2013}). 
%The additional case shown in magenta  has a larger primary MB AML. 
%magnetic braking is increased ($f_{mb}=0.3$\ in Eq.\,\ref{JMB} replaces $f_{mb}=1$\ adopted as standard case).
}
\end{center}
%\vskip -70pt
\label{figure11}       % Give a unique label
\end{figure}
As  PSR J1740--5340 belongs to a Globular Cluster, the initial mass of the MSP companion can not have been larger than the turnoff mass for this cluster. Working on this hypothesis, Burderi et al. (2002) \cite{burderi2002} reproduced the present location of the donor in the color magnitude diagram, starting the evolution with mass transfer in early case B, from a slightly evolved 0.85\msun. They presented the case that the system is now in the radio--ejection phase. Nevertheless, in the published computations the companion mass at the period of PSR J1740--5340 would be still much larger than the observed mass ($\sim 0.5$\msun, versus $\sim 0.3$\msun). 
Exploratory evolutions  include larger losses of orbital angular momentum associated with the evolution are shown in Fig.\,\ref{figure11}: the track in \cite{burderi2002} is represented by the continuous line starting at M=0.85\msun, while some  other tracks start from the stage reached at  \Porb$\sim$26\,hr. These latter are computed in the hypothesis that the NS initiates radio-ejection already at 26\,hr, it does not accrete any more and there is more or less specific angular momentum loss associated with the loss of mass from the system. Some of these tracks (full line and magenta dashed line) can indeed represent the \Mc--\Porb\ status of PSR J1740-5340, as also has been modelled by  Ergma and Sarna \cite{ergmasarna2003} who were able to fit the MSP companion with \Mc\ close to the observed range of values. 

\subsubsection{An alternative path for the short-\Porb\ MSP with He-WD companions?}
\label{hewdshort}
The peculiar evolution of binaries starting close to \Pbif\ when the mass transfer is non conservative was studied by Ergma, Sarna and Antipova \cite{ergma1998}, who  found out that when the mass is fully lost the systems evolve towards shorter \Porb, leaving an MSP plus low mass WD system. They did not specify a mechanism to account for the mass loss, but were then able to characterise a range of initial periods for which the final \Porb\  would depend on the loss of mass and AM from the system.\\
In analogy, we have shown in Fig.\,\ref{figure11} that the low donor mass and short \Porb\ of PSR\,J1740--5340 can be reproduced by including significant angular momentum losses associated with the mass lost from the binary due to radio-ejection (corresponding to the redback stage). In a similar way, the specific AML of evolutions starting from an 1.05\msun\ donor (of composition close to that of the GC Terzan\,5 ) may be adjusted  in order to reproduce the \Porb$\sim 26$\,hr of PSR J1748--2446ad (full red line in Fig.\,\ref{fig8}).

The evolution computed for PSR\,J1740--5340, but including  AML  greater than that due to mass lost with the specific AM of the donor, evolves to the very short period of about 0.5\,d (red dot--dashed line), where the mass loss stops and the companion is now a low mass He--WD. Thus the radio--ejection phase may produce remnant systems such as, discussed in Sect.\,\ref{secular}, of an MSP with a He--WD companions at  very short \Porb\ ($\sim$0.3\,d), possibly easing the difficulty mentioned by  Istrate et al. \cite{istrate2014}. \\
The MSP with He--WDs at short period may then represent an important hint of evolution with strong consequential AML due to the influence of the MSP on the mass loss from the system. Indeed, the lowest \Porb\ of such MSPs is   PSR J0751+1807 at \Porb=6.3\,hr \cite{lundgren1995}, PSR J1738+0333 at \Porb=8.51\,hr and
PSR\,J1816+4510, at \Porb=8.66\,hr \cite{kaplan2013}. In  this last system, the spectrum of the optical component resembles that of a hot ($\sim$16500K) proto-WD, but it has a lower gravity and larger radius, and it shows numerous helium and metallic lines, more similar to the non degenerate companions of redbacks. In addition, the radio MSP eclipses for $\sim$10\% of the orbit \cite{kaplan2012}, a feature again more similar to the redbacks behaviour. PSR\,J1816+4510 could represent the final stage of evolution of a binary in radio-ejection, following the stage in which it appears as a redback, and before the fully detached MSP plus He-WD stage. We show its location as a green pentagon  in the HR diagram in Fig.\,\ref{figure11}, where we see that the high \teff\ and relatively low gravity log g=5 \cite{kaplan2013} are in good agreement with the evolutionary tracks including high specific AML during the radio--ejection phase.
From this point of view, the  the transitional systems could also be linked to this peculiar evolution.

%\subsection{The redbacks and the transitional systems}
We must explicitly note that the X--ray irradiation cycles described so far have no relation whatsoever with the  `transitional' MSP, switching from the AMXP to the radio MSP on a timescale  of a few weeks \cite{papitto2015}. Such a rapid change can not be related to any of the timescales (systemic AML, thermal stellar timescale or thermal timescale of the convective envelope) so far investigated. Nevertheless, the study of irradiation shows that the thermal disequilibrium during the secular evolution is such that the stellar radius and the Roche lobe radius are kept close to each other in a very unstable situation which might well be subject to other shorter term variations. Any physical mechanism causing the transition from MSP to AMXP on weeks timescale deals with a system in which the companion barely fills its Roche lobe, in a stage of thermal instability.\\
If the redbacks are a mixed bag of systems with pre--WD or core--hydrogen burning companions, one must ask if there is a preferential class to which the transitional objects preferentially belong.

\section{Summary and (a few of the) questions left open }

In spite of all the work done in the latest 40 years, there are still problems in the binary evolution of MSP, and of their multifaceted characteristics.  We have specifically dealt mostly with the low mass -- short \Porb\ region, which is populated by very different characteristic systems, for which many questions remain open, first of all the question of next Section.

\subsection{Why the evolution close to \Pbif\ has such a dominant role?}
\label{pbifrole}
We have seen that many of the secular evolution paths of short \Porb\, -- low \Mc\  systems need to have begun close to the bifurcation period. In particular:
\begin{enumerate}
\item the radius of donors of AMXPs having \Porb\ shorter than 76\,min is the exposed core of a star which has partially depleted its initial hydrogen content (Sect.\,\ref{binevol});
\item the location of some redbacks in the \Porb--\Mc\ plane also implies that mass transfer began when the donor was substantially evolved off the MS; but not yet in the giant stage (Sect.\,\ref{scarcely}) ;
\item the presence of helium WD companions of MSP at periods 3--9\,hr  (Sect.\,\ref{hewdshort})
\item the presence of helium companions to BWs (Sect.\,\ref{ablation}).
\end{enumerate}
All these hints together point to an important role for \Pbif, both for donors of intermediate mass and for donors of low mass such as those evolving today in GCs. We leave the problem open, keeping an eye on the idea by King, Davies and Beer \cite{king2003} that some configurations are `the price of promiscuity' and are due to the evolution of MSP systems in GCs (see also Sect.\,\ref{ablation}), where binaries, in which the NS has already formed,  may suffer exchange interactions with other stars in the field of the GC, and the most suitable candidates to replace the former companion are the most massive stars alive at the epoch of the encounters. 
At least a fraction of NS formed in GCs are retained there (see Sect.\,\ref{gc1}). The fact that 40\% of MSPs are found in GCs, which include today only 10$^{-3}$ of the galactic mass, implies  that GCs are the favourite locus for accelerating the NS to MSP and for providing the best environment for the occurrence of some paths of secular evolution. Even if  the idea that the BWs in the field can be in many cases escaped from GCs \cite{king2003} may look ad hoc,  it has even been proposed that {\it all}  bulge LMXBs  may have been formed in GCs, later on destroyed by repeated tidal stripping and shocking in the galactic plane \cite{grindlayhertz1985}; further, some double neutron stars may have formed by three body interactions in GCs and ejected in the field (see Sect.\,\ref{DNS}), a relevant point to understand mergers of DNS in old stellar environments, as observed in the detection by LIGO of the event GW\,170817.     

\subsection{Possible conclusions}
We summarize here the results concerning the secular evolution of short period systems only:
\begin{enumerate}
\item  the redbacks can be mostly in a phase of radio-ejection, losing mass directly at the lagrangian point and evolving towards either shorter or longer \Porb\ according to the stage of evolution of the donor at the beginning of the mass transfer phase;

\item  the systems with He--WD companions at short \Porb\ ($<0.5$\,d) may be remnants of the evolution of the class of redbacks like PSR J740--5340, whose companion has lost and is losing mass and angular momentum due to radio--ejection. Mass exchange began after the donor developed a small helium core, and evolved off the main sequence turnoff. By losing enough specific AM, systems similar to this may evolve to shorter and shorter \Porb. Some may become `transitional' systems; 

\item  the  redback stage may begin because X--ray illumination on donors with convective envelopes occurs for a limited fraction of cycles, lasting tens to hundreds of million years, during which the system is mostly detached. Thus the companion will be subject to the MSP power in all the stages when there is no mass transfer. This is the most favourable situation to either begin the evaporation \cite{benvenuto2014} or a radio--ejection phase \cite{burderi2002}. Possibly, radio--ejection acts at longer \Porb, while evaporation is dominant for closer orbits and smaller masses. 

\item  the X--ray cycles may explain the positive large $\dot{\rm P}_{\rm orb}$\ of systems like SAX\,J1808, and the large number of MSP binaries, compared to accreting systems, both LMXB or AMXPs;

\item  evaporation and MSP illumination may be competing in dominating the evolution in the black widow stage;

%\item it is still not clear why is there a gap in the range \Mc$\sim 0.05--0.1$\msun, and why most donors in redbacks appear to have a smaller mass than expected if the evolution starts from donors which must share enough mass to accelerate the pulsar to milliseconds

\item the `transitional' stage is an open question.
\end{enumerate}

\begin{acknowledgement}
We thank Michela Mapelli, Alessandro Papitto and Enrico Vesperini for comments and suggestions.
\end{acknowledgement}
%

%\input{references}

%\bibliographystyle{spmpsci}
%\bibliography{biblo} % if your bibtex file is called example.bib

\end{document}